\documentclass[aps,prb,showpacs,tightenlines,twocolumn,secnumarabic,superscriptaddress]{revtex4-1}
\usepackage{amsmath,amssymb,amsfonts,bm}
\usepackage{graphicx}
\usepackage{epsfig}
\usepackage{epstopdf}
\usepackage{subfigure}
\usepackage{color}
\usepackage[colorlinks=true,linkcolor=blue,citecolor=blue, urlcolor=blue]{hyperref}

\begin{document}

\preprint{APS/123-QED}

\title{Instability of three-band Tomonaga-Luttinger liquid: renormalization group analysis and possible application to K$_{2}$Cr$_{3}$As$_{3}$}

\author{Jian-Jian Miao}
\affiliation{Department of Physics, Zhejiang University, Hangzhou 310027, China}
\affiliation{Collaborative Innovation Center of Advanced Microstructures, Nanjing 210093,
China}

\author{Fu-Chun Zhang}
\affiliation{Department of Physics, Zhejiang University, Hangzhou 310027, China}
\affiliation{Collaborative Innovation Center of Advanced Microstructures, Nanjing 210093,
China}

\author{Yi Zhou}
\affiliation{Department of Physics, Zhejiang University, Hangzhou 310027, China}
\affiliation{Collaborative Innovation Center of Advanced Microstructures, Nanjing 210093,
China}

\date{\today}

\begin{abstract}
Motivated by recently discovered quasi-one-dimensional superconductor K$_{2}$Cr$_{3}$As$_{3}$ with $D_{3h}$ lattice symmetry, we study one-dimensional
three-orbital Hubbard model with generic electron repulsive interaction described by intra-orbital repulsion $U$, inter-orbital repulsion $U'$, and Hund's coupling $J$.
As extracted from density functional theory calculation, two of the three atomic orbitals are degenerate ($E^{\prime}$ states) and the third one is non-degenerate ($A^{\prime}_1$),
and the system is at incommensurate filling. With the help of bosonization, the normal state is described by a three-band Tomonaga-Luttinger liquid.
Possible charge density wave (CDW), spin density wave (SDW) and superconducting (SC) instabilities are analyzed by renormalization group method.
The ground state depends on the ratio $J/U$ and is sensitive to the degeneracy of $E^{\prime}$ bands.  Spin-singlet SC state is favored at $0<J<U/3$, and spin-triplet SC state is favored in the region $U/3<J<U/2$.
The SDW state has the lowest energy only in the unphysical parameter region $J>U/2$.  When the two-fold degeneracy of $E^{\prime}$ bands is lifted,   SDW instability has the tendency to dominate over the spin-singlet SC state at $0<J<U/3$, while the order parameter of the spin-triplet SC state will be modulated by a phase factor $2\Delta k_F x$ at $U/3<J<U/2$. Possible experimental consequences and applications to K$_{2}$Cr$_{3}$As$_{3}$ are discussed.
\end{abstract}

\maketitle


\section{Introduction}

There has been considerable interest on the recent discovery of a new family of quasi-one-dimensional (quasi-1D) unconventional superconductors A$_{2}$Cr$_{3}$As$_{3}$(A = K, Rb, Cs) in ambient pressure with $T_c$ up to 6.1 K \cite{GHCao_K,GHCao_Rb,GHCao_Cs},
because of their exotic properties revealed in various experiments below.
(1) In the normal state, the resistivity in polycrystalline samples follows a linear temperature dependence, $\rho (T)=\rho _{0}+AT$, in a wide temperature region, different from the usual Fermi liquid behavior $\rho _{0}+AT^{2}$\cite{GHCao_K,GHCao_Rb,GHCao_Cs}.
On the other hand, the transport measurement in single crystalline samples indicates that the normal state is a smectic metal, namely, it behaves as a metal along the $c$-axis and a semiconductor in the $ab$-plane \cite{GHCaoabc}.
(2) Nuclear magnetic resonance (NMR) and nuclear quadrupole resonance (NQR) measurements on K$_{2}$Cr$_{3}$As$_{3}$ show a non-integer power-law temperature dependence $1/T_{1}\sim T^{0.75} $ above $T_{c}$,
which is neither $1/T_{1}\sim T$ for a Fermi liquid nor Curie-Weiss behavior $1/T_{1}T\sim C/(T+\theta )$ for a ferromagnet or antiferromagnet \cite{TImai_K}.
Meanwhile, NMR and NQR experiments on Rb$_{2}$Cr$_{3}$As$_{3}$ show a critical spin fluctuation above $T_c$, $1/T_{1}T\sim a + b/(T+\theta )$, where $\theta\sim 0$K \cite{GQZheng_Rb}.
The Hebel-Slichter coherence peak of $1/T_{1}$ is absent in both compounds.
(3) K$_{2}$Cr$_{3}$As$_{3}$ possesses a large upper critical field $H_{c2}$, which exceeds the BCS weak-coupling Pauli limit field by 3-4 times \cite{GHCao_K,PCCanfield_K,RDMcDonald_K}.
The angle resolved $H_{c2}$ measurement demonstrates strong anisotropy and reveals dominant spin-triplet SC pairing \cite{ZWZhu_K},
which is consistent with the observation of a very weak spontaneous internal magnetic field near $T_c$ in the muon spin relaxation/rotation ($\mu$SR) experiment \cite{muSR_K}.
(4) London penetration depth measurement for K$_{2}$Cr$_{3}$As$_{3}$ shows linear temperature dependence, $\Delta\lambda (T)\sim T$, at temperatures $T\ll T_{c}$, indicating the existence of line nodes in the SC gap \cite{HQYuan_K}.
(5) Doping nonmagnetic impurities in K$_{2}$Cr$_{3}$As$_{3}$ will reduce $T_c$ significantly, which indicates non-$s$-wave superconductivity \cite{GHCao_impurity}.

There have also been a series of theoretical studies. (1) The electronic structure of K$_{2}$Cr$_{3}$As$_{3}$ has been investigated by Jiang \textit{et al.}\cite{CCao_K} using density functional theory (DFT),
which is confirmed by later calculation \cite{JPHu_magnetism}. The band calculations show that Cr-3$d$ orbitals dominate the electronic states near the Fermi level,
and there exist three energy bands at the Fermi level: two quasi-1D $\alpha $- and $\beta $-bands with flat Fermi surfaces, and a 3D $\gamma $-band.
(2) Zhou et al. \textit{et al.} proposed a minimum effective model based on three molecular orbitals on a hexagonal lattice with $D_{3h}$ symmetry \cite{YZhou_threeband}.
They found that for small Hubbard $U$ and moderate Hund's coupling $J$, the pairing arises from the 3D $\gamma$ band and has a spatial symmetry $f_{y\left(3x^{2}-y^{2}\right)}$, which gives line nodes in the gap function,
while for large $U$, a fully gapped $p$-wave state, $p_{z}\hat{z}$ dominates at the quasi-1D $\alpha$-band. The spin-triplet SC pairing is driven by the Hund's coupling.
Similar three-band and six-band models were also proposed by Wu \textit{et al.} \cite{JPHu_threeband,JPHu_sixband,JPHu_experiment}.
The dominant SC instability channels are found as $p_z$ and $f_{y\left(3x^{2}-y^{2}\right)}$ for weak and strong Hund's coupling respectively.
(3) Zhong \textit{et al.} carried out DFT calculation on a single [CrAs]$_{\infty}$ tube to construct an effective three-band Hubbard model \cite{JHDai_TLL}. Possible Tomonaga-Luttinger liquid instabilities have been proposed based on such a three-band Hubbard chain.

Besides its possible exotic superconductivity, K$_{2}$Cr$_{3}$As$_{3}$ provides a platform for us to study 1D correlated electrons apart from carbon nanotubes and cuprate ladders.
The key building block of K$_{2}$Cr$_{3}$As$_{3}$ is the 1D [(Cr$_3$As$_3$)$^{2-}$]$_{\infty}$ double-walled subnanotubes, which are separated by columns of K$^{+}$ ions, in contrast to the layered iron-pnictide and
copper-oxide high Tc superconductors \cite{GHCao_K}. These [(Cr$_3$As$_3$)$^{2-}$]$_{\infty}$ tubes together with K$^{+}$ ions form a noncentrosymmetric hexagonal lattice with $D_{3h}$ point group \cite{GHCao_K}.
The quasi-one-dimensionality can be also seen from its electronic structure, say, the existence of two quasi-1D electron bands \cite{CCao_K,JPHu_magnetism}.
Experimentally, both the smectic metallic transport \cite{GHCaoabc} and the non-integer power-law temperature dependence in NMR $1/T_{1}\sim T^{0.75}$ \cite{TImai_K} imply a Tomonaga-Luttinger liquid (TLL) normal state.
The question is how this three-band TLL normal state gives rise to the unconventional SC states below $T_c$. This motivates us to study possible instabilities of three-band TLLs in this paper. Our analysis on the TLLs in this class of materials may also help us understand their normal state properties.

It is noted that two-leg Hubbard ladders and two-orbital Hubbard chains have already been investigated \cite{Chudzinski,AJMillis_twoband}, and that the
SC instability caused by electron-phonon coupling in three-band metallic nanotubes has also been theoretically studied \cite{EOrignac_threeband}.
In this paper, we shall focus on electron interactions in a 1D three-band Hubbard model.

This paper is organized as follows. We present the electronic model Hamiltonian in Section~\ref{model}.
In Section~\ref{g-ology}, the low-energy scattering processes near the Fermi points are classified by using generalized $g$-ology.
In Section~\ref{bosonization}, we take the continuum limit and use bosonization technique to transform the fermionic Hamiltonian into bosonic Hamiltonian. The non-interacting part describes a three-band TLL, and
the remaining terms describe the bosonic interactions.
In Section~\ref{order}, order parameters are defined to characterize ordered states.
In Section~\ref{RG}, we utilize renormalization group (RG) to analyze these bosonic interactions. The RG equations are derived by operator product expansion (OPE) method.
The relevant terms lead to different instabilities in different parameters regions.
Section~\ref{conclusion} is devoted to discussions and conclusions.

\section{model Hamiltonian}\label{model}

We consider a single fermionic chain with a unit cell (per Cr$_6$As$_6$ cluster) containing three molecular orbitals.
One of the three orbitals belongs to one-dimensional irreducible representation $A_{1}^{\prime}$ of $D_{3h}$ group, and the other two are in the two-dimensional irreducible representation $E^{\prime}$ \cite{YZhou_threeband}.
Without loss of generality, the fermionic Hamiltonian consists of two parts,
\begin{subequations}
\begin{equation}
\begin{aligned}
H^{F}=H_{0}^{F}+H_{int}^{F},
\end{aligned}
\end{equation}
where the non-interacting part $H_{0}^{F}$ is a three-band tight-binding Hamiltonian describing the electron hopping, while the interacting part $H_{int}^{F}$ originates from the electron-electron interaction.

The $D_{3h}$ lattice symmetry does not allow the mixture between $A_1^{\prime}$ state and $E^{\prime}$ states along the $c$-direction. The absence of such hybridization is also seen from the DFT calculation,
where the $\beta$ and $\gamma$ bands are degenerate along the $\Gamma-A$ line.
Neglecting the inter-chain coupling, we have the following $H^F_0$ in such a 1D system,
\begin{equation}\label{Eq:tight-binding}
\begin{aligned}
H_{0}^{F}=\sum_{km\sigma}\xi_{km}c_{km\sigma}^{\dagger}c_{km\sigma},
\end{aligned}
\end{equation}
where $\sigma=\uparrow,\downarrow$ is the spin index, and the orbital (or band) index $m=0$ refers to the $A_1^{\prime}$ state and $m=\pm 1$ refer to $E^{\prime}$ states.
$c_{km\sigma}$($c_{km\sigma}^{\dagger}$) is electron annihilation (creation) operator for orbital $m$ and spin $\sigma$. 
The band structure from tight-binding model\cite{YZhou_threeband} is plotted in Fig.~\ref{fig:dispersion}, where the linearized energy dispersion near the Fermi energy is shown in the inset.

\begin{figure}[hptb]
\includegraphics[width=9.2cm]{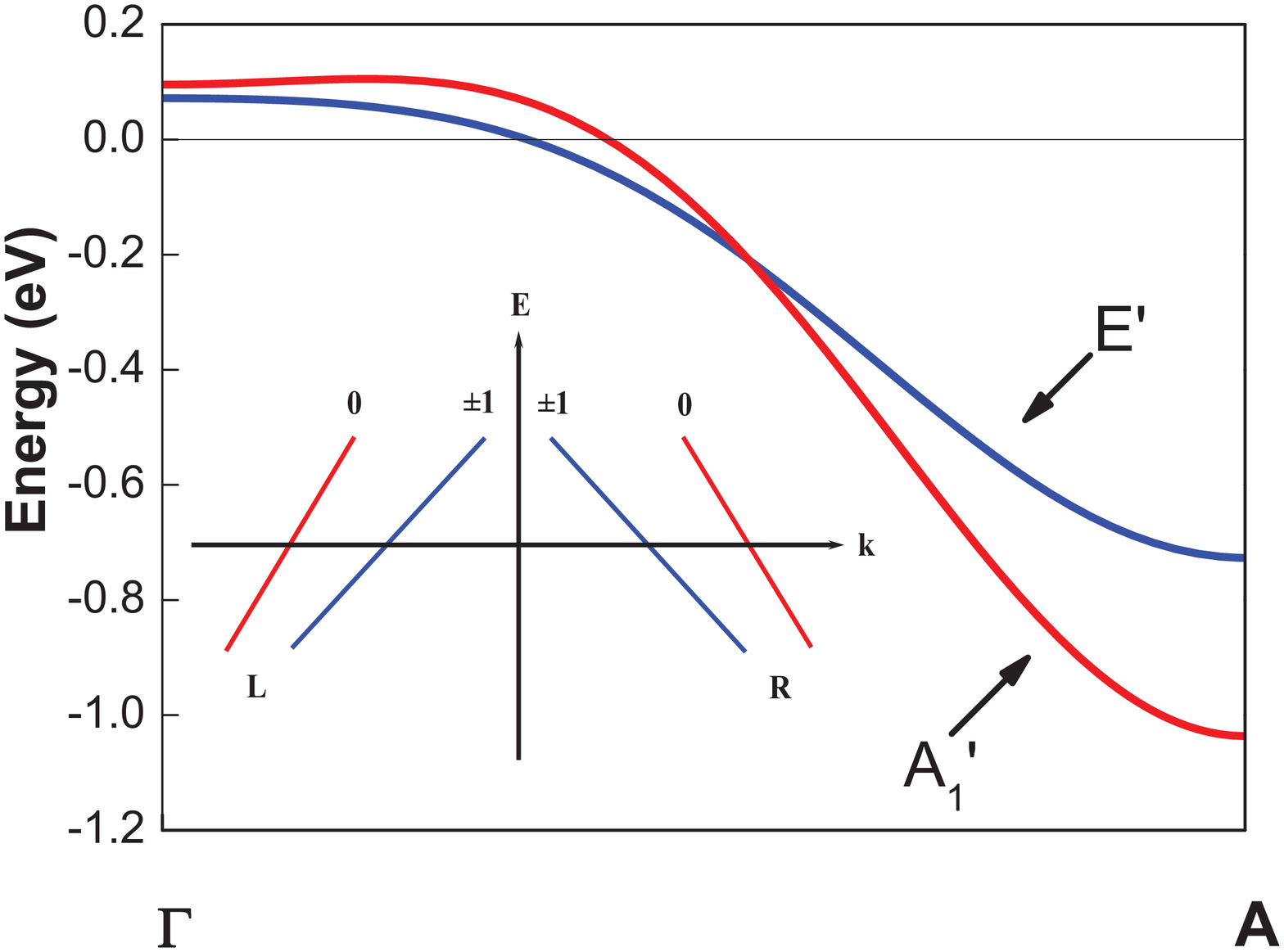}
\caption {Band structure from tight-binding model. The $A_1^{\prime}$ band is nondegenerate and the $E^{\prime}$ band is two-fold degenerate. $\Gamma=(0,0,0)$ and $A=(0,0,\pi)$ in the reciprocal space. 
Inset shows the linearized energy dispersion near the Fermi energy.}
\label{fig:dispersion}
\end{figure}

The interacting part $H_{int}^{F}$ describes electron interactions. In the Hubbard approximation, we only retain on-site Coulomb repulsion. The interaction Hamiltonian contains four terms
\begin{eqnarray}\label{Eq:Hubbard}
H_{int}^{F} & = & \frac{1}{2}\sum_{im}\sum_{\sigma\neq\sigma'}Un_{im\sigma}n_{im\sigma'}+\frac{1}{2}\sum_{i\sigma\sigma'}\sum_{m\neq m'}U'n_{im\sigma}n_{im'\sigma'}\nonumber \\
 & - & \sum_{i}\sum_{m\neq m'}J\left(\vec{S}_{im}\cdot\vec{S}_{im'}+\frac{1}{4}n_{im}n_{im'}\right)\nonumber \\
 & + & \frac{1}{2}\sum_{i\sigma}\sum_{m\neq m'}J'c_{im\sigma}^{\dagger}c_{im\bar{\sigma}}^{\dagger}c_{im'\bar{\sigma}}c_{im'\sigma},
\end{eqnarray}
\end{subequations}
where $n_{im\sigma}=c_{im\sigma}^{\dagger}c_{im\sigma}$, $n_{im}=\sum_{\sigma}n_{im\sigma}$, $\vec{S}_{im}=\frac{1}{2}\sum_{\alpha\beta}c_{im\alpha}^{\dagger}\vec{\tau}_{\alpha\beta}c_{im\beta}$,
$\vec{\tau}$ is a vector with three components of Pauli matrices, and $\bar{\sigma}=-\sigma$ is the opposite spin to $\sigma$.
$U$ is the intra-orbital repulsion, $U'$ is the inter-orbital repulsion, $J$ is the Hund's coupling, and $J'$ is the pair-hopping.
Note that we have chosen Wannier functions to be real. The two degenerate orbitals $m=\pm1$ transfer as $x$ and $y$ under $D_{3h}$ symmetry operations, respectively.
We also assume that
\begin{equation}\label{Eq:J}
J'=J>0,
\end{equation}
so that the following relation
\begin{equation}\label{Eq:U-J}
U=U'+2J
\end{equation}
arises subject to the rotational symmetry of the Coulomb interaction.

It is noted that similiar models for three coupled chains \cite{EArrigoni_threechain} and three-leg ladders \cite{TKimura_threeleg} have been investigated using renormalization group. 
The important difference between these existing models and present model is that two of the three bands are degenerate or nearly degenerate in our case, which plays a crucial role for superconducting instabilities as we will see in next sections.

\section{Continuum limit and the $g$-ology}\label{g-ology}

Now we introduce electron fields $c_{m\sigma}(x)$ to study the low energy physics in the continuum limit, hereafter $x$ denotes the coordinate along the chain ($c$-direction).
In a 1D system, Fermi points fall into two categories characterized by chirality $p=R,L$, which represents right and left-moving electrons, respectively. Thus the electron field $c_{m\sigma}(x)$ can be decomposed into two parts
\begin{equation}\label{Eq:c-field}
c_{m\sigma}(x)=\psi_{Rm\sigma}(x)+\psi_{Lm\sigma}(x)
\end{equation}
in low energies.

\begin{figure}[hptb]
\begin{center}
\includegraphics[width=8.4cm]{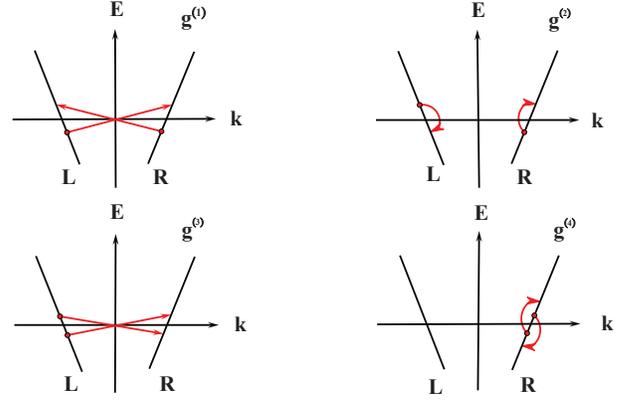}
\end{center}
\caption{
Four possible scattering processes for signle-band spinless fermions:
$g^{(1)}\psi_{p}^{\dagger}\psi_{\bar{p}}^{\dagger}\psi_{p}\psi_{\bar{p}}$,
$g^{(2)}\psi_{p}^{\dagger}\psi_{\bar{p}}^{\dagger}\psi_{\bar{p}}\psi_{p}$, $g^{(3)}\psi_{p}^{\dagger}\psi_{p}^{\dagger}\psi_{\bar{p}}\psi_{\bar{p}}$
and $g^{(4)}\psi_{p}^{\dagger}\psi_{p}^{\dagger}\psi_{p}\psi_{p}$, where $\bar{p}$ is the opposite chirality to $p$.}
\label{fig:g-ology}
\end{figure}

In order to classify various scattering processes in such a three-band system, we shall generalize the conventional $g$-ology \cite{TGiamarchi_bosonization,AJMillis_twoband} for single-band spinless fermions, which now includes chirality, band and spin indices.
For single-band spinless fermions, there are four possible scattering processes between the two chiralities because of lattice momentum conservation.
All these scattering processes are illustrated in Fig.~\ref{fig:g-ology},
back scattering $g^{(1)}\psi_{p}^{\dagger}\psi_{\bar{p}}^{\dagger}\psi_{p}\psi_{\bar{p}}$,
double-chirality forward scattering $g^{(2)}\psi_{p}^{\dagger}\psi_{\bar{p}}^{\dagger}\psi_{\bar{p}}\psi_{p}$,
umklapp scattering $g^{(3)}\psi_{p}^{\dagger}\psi_{p}^{\dagger}\psi_{\bar{p}}\psi_{\bar{p}}$ and single-chirality forward scattering $g^{(4)}\psi_{p}^{\dagger}\psi_{p}^{\dagger}\psi_{p}\psi_{p}$,
where $\bar{p}$ is the opposite chirality to $p$.

\begin{table*}[htpb]
\caption{$g$-ology for the three-band spinful fermion system.}
\label{table:g-gology}
\begin{center}
\begin{tabular}{|c|l|c|c|c|c|c|l|}
\hline
\multicolumn{1}{|c}{} & chirality & \multicolumn{1}{c}{} & \multicolumn{1}{c}{} & \multicolumn{1}{c}{band} &  & \multicolumn{1}{c}{} & spin\tabularnewline
\hline
$g(f)^{(1)}$ & $\psi_{p}^{\dagger}\psi_{\bar{p}}^{\dagger}\psi_{p}\psi_{\bar{p}}$ & $g_{1}$ & $\psi_{m}^{\dagger}\psi_{\bar{m}}^{\dagger}\psi_{m}\psi_{\bar{m}}$ & $f_{1}$ & $\psi_{m}^{\dagger}\psi_{0}^{\dagger}\psi_{m}\psi_{0}+h.c.$ & $g(f)_{\parallel}$ & $\psi_{\sigma}^{\dagger}\psi_{\sigma}^{\dagger}\psi_{\sigma}\psi_{\sigma}$\tabularnewline
\hline
$g(f)^{(2)}$ & $\psi_{p}^{\dagger}\psi_{\bar{p}}^{\dagger}\psi_{\bar{p}}\psi_{p}$ & $g_{2}$ & $\psi_{m}^{\dagger}\psi_{\bar{m}}^{\dagger}\psi_{\bar{m}}\psi_{m}$ & $f_{2}$ & $\psi_{m}^{\dagger}\psi_{0}^{\dagger}\psi_{0}\psi_{m}+h.c.$ & $g(f)_{\perp}$ & $\psi_{\sigma}^{\dagger}\psi_{\bar{\sigma}}^{\dagger}\psi_{\bar{\sigma}}\psi_{\sigma}$\tabularnewline
\hline
$g(f)^{(3)}$ & $\psi_{p}^{\dagger}\psi_{p}^{\dagger}\psi_{\bar{p}}\psi_{\bar{p}}$ & $g_{3}$ & $\psi_{m}^{\dagger}\psi_{m}^{\dagger}\psi_{\bar{m}}\psi_{\bar{m}}$ & $f_{3}$ & $\psi_{m}^{\dagger}\psi_{m}^{\dagger}\psi_{0}\psi_{0}+h.c.$ &  & \tabularnewline
\hline
$g(f)^{(4)}$ & $\psi_{p}^{\dagger}\psi_{p}^{\dagger}\psi_{p}\psi_{p}$ & $g_{4}$ & $\psi_{m}^{\dagger}\psi_{m}^{\dagger}\psi_{m}\psi_{m}$ & $g$ & $\psi_{0}^{\dagger}\psi_{0}^{\dagger}\psi_{0}\psi_{0}$ &  & \tabularnewline
\hline
\end{tabular}
\end{center}
\end{table*}

\begin{figure*}[hptb]
\begin{center}
\subfigure[~$g_{1\perp}^{(2)}\psi_{pm\sigma}^{\dagger}\psi_{\bar{p}\bar{m}\bar{\sigma}}^{\dagger}\psi_{\bar{p}m\bar{\sigma}}\psi_{p\bar{m}\sigma}$]{\includegraphics[width=6.4cm]{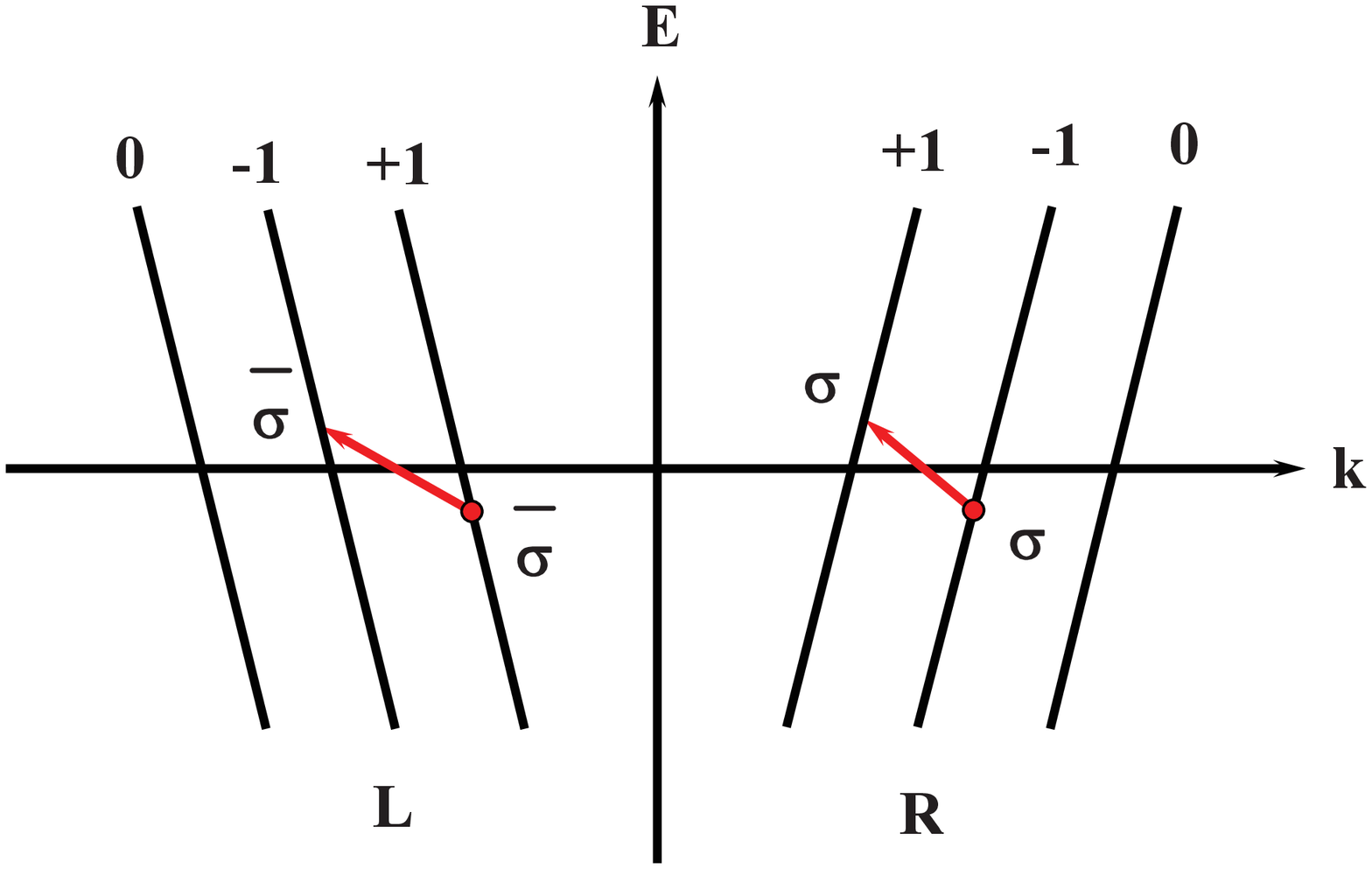}}
\subfigure[~$g_{2\parallel}^{(1)}\psi_{pm\sigma}^{\dagger}\psi_{\bar{p}\bar{m}\sigma}^{\dagger}\psi_{p\bar{m}\sigma}\psi_{\bar{p}m\sigma}$]{\includegraphics[width=6.4cm]{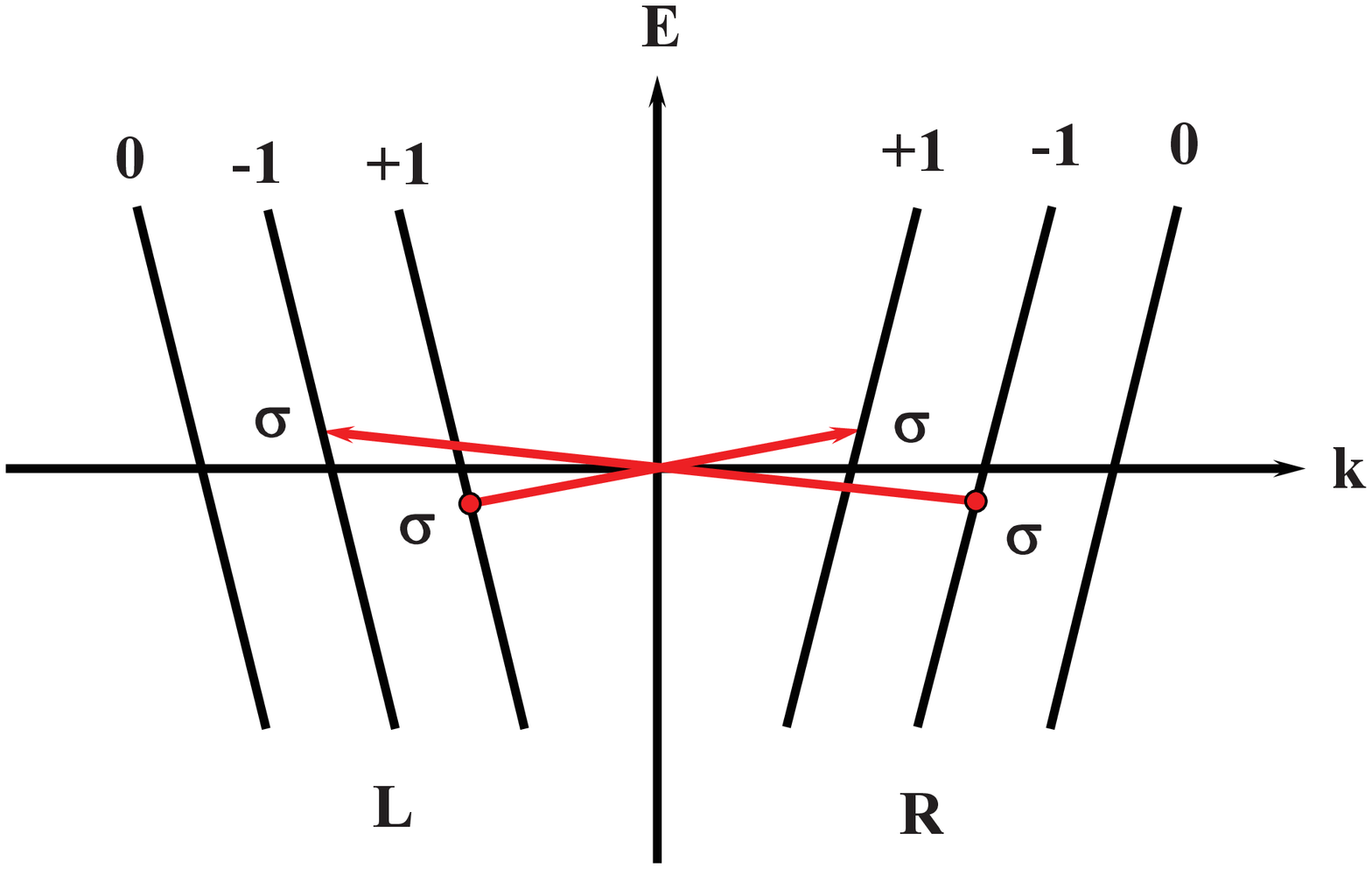}}\\
\subfigure[~$g_{3\parallel}^{(1)}\psi_{pm\sigma}^{\dagger}\psi_{\bar{p}m\sigma}^{\dagger}\psi_{p\bar{m}\sigma}\psi_{\bar{p}\bar{m}\sigma}$]{\includegraphics[width=6.4cm]{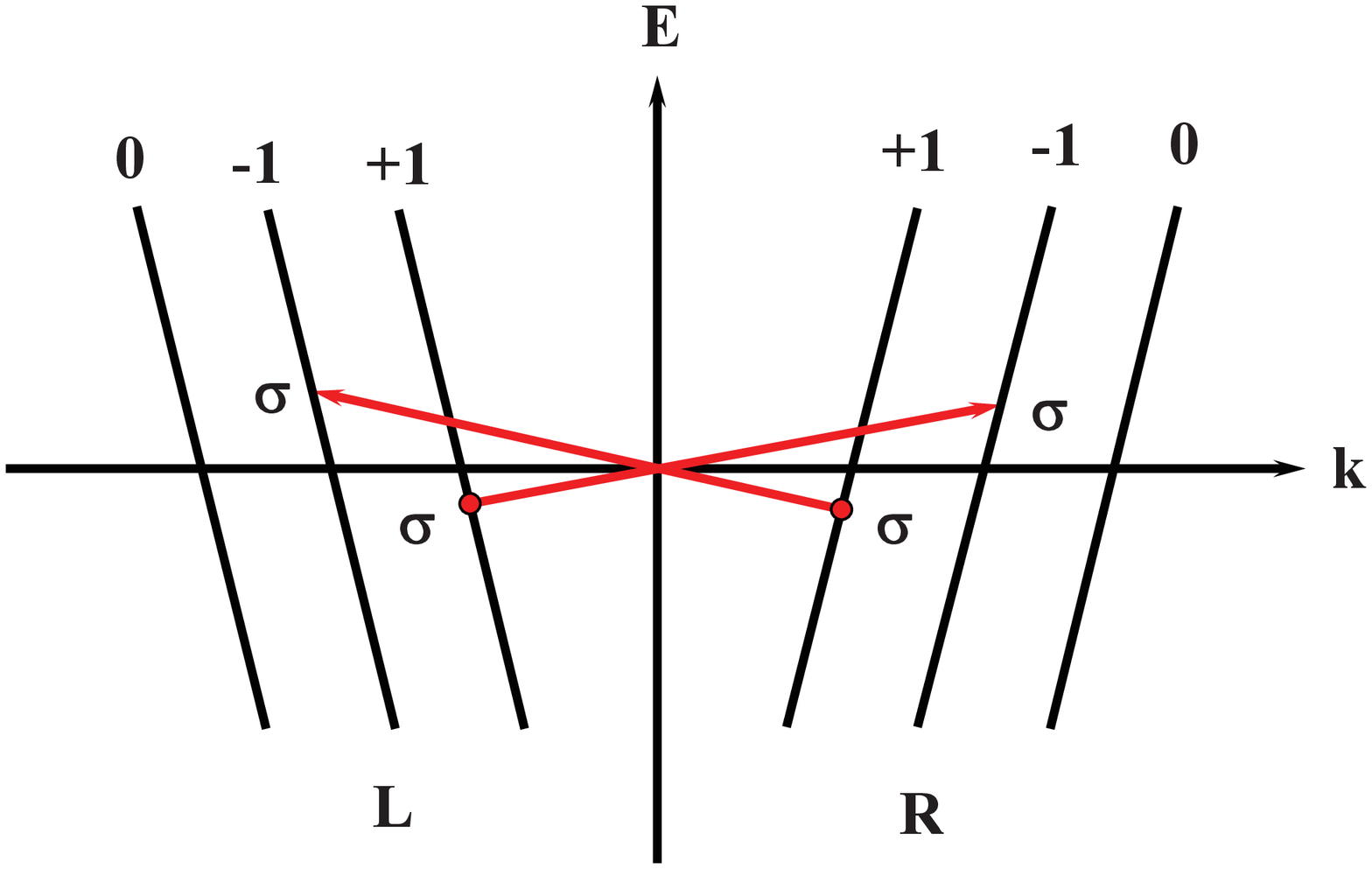}}
\subfigure[~$f_{3\parallel}^{(1)}\psi_{pm\sigma}^{\dagger}\psi_{\bar{p}m\sigma}^{\dagger}\psi_{p0\sigma}\psi_{\bar{p}0\sigma}$]{\includegraphics[width=6.4cm]{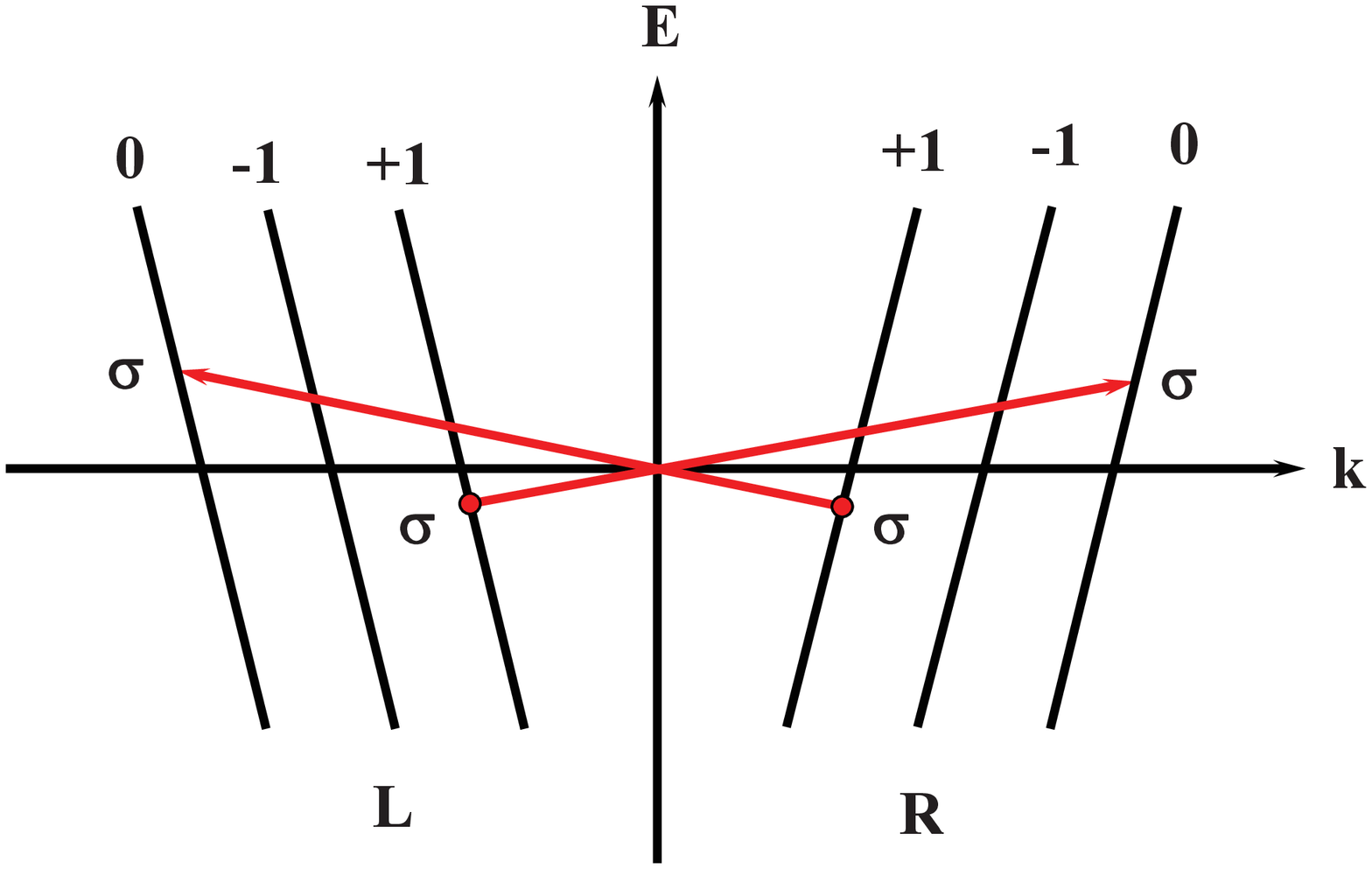}}
\end{center}
\caption{ Four dominant scattering processes at incommensurate filling: (a) $g_{1\perp}^{(2)}\psi_{pm\sigma}^{\dagger}\psi_{\bar{p}\bar{m}\bar{\sigma}}^{\dagger}\psi_{\bar{p}m\bar{\sigma}}\psi_{p\bar{m}\sigma}$,
(b) $g_{2\parallel}^{(1)}\psi_{pm\sigma}^{\dagger}\psi_{\bar{p}\bar{m}\sigma}^{\dagger}\psi_{p\bar{m}\sigma}\psi_{\bar{p}m\sigma}$,
(c) $g_{3\parallel}^{(1)}\psi_{pm\sigma}^{\dagger}\psi_{\bar{p}m\sigma}^{\dagger}\psi_{p\bar{m}\sigma}\psi_{\bar{p}\bar{m}\sigma}$
and (d) $f_{3\parallel}^{(1)}\psi_{pm\sigma}^{\dagger}\psi_{\bar{p}m\sigma}^{\dagger}\psi_{p0\sigma}\psi_{\bar{p}0\sigma}$.}
\label{fig:example}
\end{figure*}

For the three-band spinful fermions, we introduce an additional notation $f$ and additional two subscripts to describe the scattering processes due to the multi-bands and spin degrees of freedom, which is summarized in Table \ref{table:g-gology}.
One of the subscripts is for spin degrees of freedom, namely, ``$\parallel$" denotes spin parallel scattering and ``$\perp$" denotes spin anti-parallel scattering.
The other subscript is associated with the notations $g$ and $f$.
Now the notation $g$ is used only for the scattering processes within the same $D_{3h}$ irreducible representation, which includes the scattering between two $E^{\prime}$ bands with $m=\pm 1$ and the scattering within the $A_1^{\prime}$ band with $m=0$.
It is similar to $g^{1,2,3,4}$ for two chiralities that we use $g_{1} \psi_{m}^{\dagger}\psi_{\bar{m}}^{\dagger}\psi_{m}\psi_{\bar{m}}$,
$g_{2} \psi_{m}^{\dagger}\psi_{\bar{m}}^{\dagger}\psi_{\bar{m}}\psi_{m}$,
$g_{3} \psi_{m}^{\dagger}\psi_{m}^{\dagger}\psi_{\bar{m}}\psi_{\bar{m}}$,
and $g_{4} \psi_{m}^{\dagger}\psi_{m}^{\dagger}\psi_{m}\psi_{m}$ for the scatterings between two $E^{\prime}$ bands, where $\bar{m}$ is the opposite orbital to $m$.
We also use $g \psi_{0}^{\dagger}\psi_{0}^{\dagger}\psi_{0}\psi_{0}$ for the scattering within the $A_1^{\prime}$ band by neglecting the subscript.
On the other hand, the new notation $f$ describes the scattering between $E^{\prime}$ and $A^{\prime}_1$ bands, including
$f_{1}(\psi_{m}^{\dagger}\psi_{0}^{\dagger}\psi_{m}\psi_{0}+h.c.)$, $f_{2}(\psi_{m}^{\dagger}\psi_{0}^{\dagger}\psi_{0}\psi_{m}+h.c.)$, and $f_{3}(\psi_{m}^{\dagger}\psi_{m}^{\dagger}\psi_{0}\psi_{0}+h.c.)$, where $m=\pm 1$.
Four typical scattering processes are plotted in Fig.~\ref{fig:example}, which are all the dominant scattering processes at incommensurate filling as we will discuss later.

The long wavelength physics is dominated by low energy scattering processes near the Fermi points.
These $g$-ology classified scattering processes serve as building blocks for the low energy effective theory. We can decouple the microscopic Hamiltonian in terms of these processes to obtain the effective theory.

For instance, the inter-band Hubbard repulsive interaction $U'$ between the two degenerate $E^{\prime}$ bands $m=\pm1$ can be decoupled as follows,
\begin{eqnarray}
 &  & U'\sum_{m\sigma\sigma'}c_{im\sigma}^{\dagger}c_{im\sigma}c_{i\bar{m}\sigma'}^{\dagger}c_{i\bar{m}\sigma'}\nonumber \\
 & = & U'\sum_{pm\sigma}\left(\psi_{pm\sigma}^{\dagger}\psi_{\bar{p}\bar{m}\sigma}^{\dagger}\psi_{p\bar{m}\sigma}\psi_{\bar{p}m\sigma} + \psi_{pm\sigma}^{\dagger}\psi_{\bar{p}\bar{m}\sigma}^{\dagger}\psi_{\bar{p}\bar{m}\sigma}\psi_{pm\sigma}\right.\nonumber\\
 &  &  + \psi_{pm\sigma}^{\dagger}\psi_{p\bar{m}\sigma}^{\dagger}\psi_{\bar{p}\bar{m}\sigma}\psi_{\bar{p}m\sigma} + \psi_{pm\sigma}^{\dagger}\psi_{p\bar{m}\sigma}^{\dagger}\psi_{p\bar{m}\sigma}\psi_{pm\sigma}\nonumber \\
 &  &  + \psi_{pm\sigma}^{\dagger}\psi_{\bar{p}\bar{m}\bar{\sigma}}^{\dagger}\psi_{p\bar{m}\bar{\sigma}}\psi_{\bar{p}m\sigma} + \psi_{pm\sigma}^{\dagger}\psi_{\bar{p}\bar{m}\bar{\sigma}}^{\dagger}\psi_{\bar{p}\bar{m}\bar{\sigma}}\psi_{pm\sigma}\nonumber\\
 &  &  \left. + \psi_{pm\sigma}^{\dagger}\psi_{p\bar{m}\bar{\sigma}}^{\dagger}\psi_{\bar{p}\bar{m}\bar{\sigma}}\psi_{\bar{p}m\sigma} + \psi_{pm\sigma}^{\dagger}\psi_{p\bar{m}\bar{\sigma}}^{\dagger}\psi_{p\bar{m}\bar{\sigma}}\psi_{pm\sigma}\right).
\end{eqnarray}

The initial values of the coupling constants ($f$'s and $g$'s) in the effective theory are determined by the microscopic Hamiltonian $H^{F}$.
Decoupling all the terms in $H^{F}_{int}$ in Eq.~(\ref{Eq:Hubbard}) and collecting all the scattering processes, we obtain the values of nonzero coupling constants,
\begin{subequations}\label{Eq:couplings}
\begin{align}
g_{1\perp}^{(1)}&=g_{1\perp}^{(2)}=g_{3\perp}^{(1)}=g_{3\perp}^{(2)}\nonumber\\
                           &=f_{1\perp}^{(1)}=f_{1\perp}^{(2)}=f_{3\perp}^{(1)}=f_{3\perp}^{(2)}=J,      \\
g_{4\perp}^{(1)}&=g_{4\perp}^{(2)}=g_{\perp}^{(1)}=g_{\perp}^{(2)}=U, \\
g_{2\perp}^{(1)}&=g_{2\perp}^{(2)}=f_{2\perp}^{(1)}=f_{2\perp}^{(2)}=U-2J, \\
g_{2\parallel}^{(1)}&=g_{2\parallel}^{(2)}=f_{2\parallel}^{(1)}=f_{2\parallel}^{(2)}=U-3J,
\end{align}
\end{subequations}
where the relations Eq.~(\ref{Eq:J}) and Eq.~(\ref{Eq:U-J}) have been used in deriving the above equations. Finally, we shall take the continuum limit
by using Eq.~(\ref{Eq:c-field}) to obtain the fermion field theory.

\section{bosonization}\label{bosonization}

To study low energy effective theory, we shall utilize the standard bosonization technique to analyze the continuum fermion model.
In abelian bosonization, the fermion operators can be expressed in terms of boson operators as follows \cite{TGiamarchi_bosonization},
\begin{subequations}\label{Eq:bosonization}
\begin{equation}
\psi_{pm\sigma}=\frac{\eta_{m\sigma}}{\sqrt{2\pi a}}e^{ipk_{Fm}x}e^{-ip\varphi_{pm\sigma}},
\end{equation}
where $k_{Fm}$ is the Fermi momentum for band $m$, $a$ is the cutoff which can be chosen as the lattice constant, and $p=1(-1)$ stands for $R(L)$ branch.
The Klein factors $\eta_{m\sigma}$ ensure the fermionic statistics and obey the anticommutation relations
\begin{equation}
\left\{ \eta_{m\sigma},\eta_{m'\sigma'}\right\} =2\delta_{mm'}\delta_{\sigma\sigma'}.
\end{equation}
Counting the four-fermion interactions, there are still some gauge degrees of freedom for choosing the values of product of two Klein factors with different band indices $m$.
In this paper, we adopt the convention
\begin{align}
\eta_{m\sigma}\eta_{\bar{m}\sigma}&=\eta_{0\sigma}\eta_{m\sigma}=im\sigma, \\
\eta_{m\sigma}\eta_{m\bar{\sigma}}&=\eta_{0\sigma}\eta_{0\bar{\sigma}}=i\sigma, \\
\eta_{m\sigma}\eta_{\bar{m}\bar{\sigma}}&=\eta_{0\sigma}\eta_{m\bar{\sigma}}=im,
\end{align}
where $m=\pm1$ and $\sigma=+1\left(-1\right)$ for spin up(down). As we will see later in this section, these products of two Klein factors will determine the sign of coupling constants in the bosonic interacting Hamiltonian.

The chiral fields $\varphi_{pm\sigma}$ can be written in terms of two non-chiral fields $\phi_{m\sigma}$ and $\theta_{m\sigma}$ through
\begin{equation}
\varphi_{pm\sigma}=\phi_{m\sigma}-p\theta_{m\sigma}.
\end{equation}
Their gradients are proportional to fermionic density and current operator respectively,
\begin{align}
\nabla\phi_{m\sigma}\varpropto n_{m\sigma}&=\psi_{Rm\sigma}^{\dagger}\psi_{Rm\sigma}+\psi_{Lm\sigma}^{\dagger}\psi_{Lm\sigma}, \\
\nabla\theta_{m\sigma}\varpropto j_{m\sigma}&=\psi_{Rm\sigma}^{\dagger}\psi_{Rm\sigma}-\psi_{Lm\sigma}^{\dagger}\psi_{Lm\sigma}.
\end{align}
Thus the four-fermion density-density and current-current interaction can be bosonized into quadratic terms in the bosonic Hamiltonian.

Furthermore, the fields $\phi_{m\sigma}$ and $\theta_{m\sigma}$ can be decomposed into their charge and spin degrees of freedom,
\begin{align}
\phi_{m\sigma}=\frac{1}{\sqrt{2}}\left(\phi_{cm}+\sigma\phi_{sm}\right), \\
\theta_{m\sigma}=\frac{1}{\sqrt{2}}\left(\theta_{cm}+\sigma\theta_{sm}\right).
\end{align}
\end{subequations}
Since both charge and spin are conserved, $\phi_{cm}(\theta_{cm})$ and $\phi_{sm}(\theta_{sm})$ can be diagonalized separately in the quadratic part of the bosonic Hamiltonian $H_0^B$.
The diagonalization can be carried out explicitly by the following transformation,
\begin{equation}\label{Eq:diagonalize-HB0}
\left(\begin{array}{c}
\phi(\theta)_{\mu+1}\\
\phi(\theta)_{\mu-1}\\
\phi(\theta)_{\mu0}
\end{array}\right)=\left(\begin{array}{ccc}
\frac{1}{\sqrt{2}} & \frac{1}{\sqrt{6}} & \frac{1}{\sqrt{3}}\\
-\frac{1}{\sqrt{2}} & \frac{1}{\sqrt{6}} & \frac{1}{\sqrt{3}}\\
0 & -\frac{2}{\sqrt{6}} & \frac{1}{\sqrt{3}}
\end{array}\right)\left(\begin{array}{c}
\tilde{\phi}(\tilde{\theta})_{\mu+1}\\
\tilde{\phi}(\tilde{\theta})_{\mu-1}\\
\tilde{\phi}(\tilde{\theta})_{\mu0}
\end{array}\right),
\end{equation}
where $\mu=c,s$ refers to charge and spin components.

Near the Fermi points, the energy dispersion $\xi_{km}$ can be linearized as
\begin{equation}
\xi_{km}=v_{Fm}\left(k-k_{Fm}\right),
\end{equation}
where $v_{Fm}$ is the Fermi velocity and $k_{Fm}$ is the Fermi momentum. According to the DFT calculation, the difference between $v_{F0}$ and $v_{F\pm1}$ is small.
As we will show below,  the Fermi velocity is renormalized by the forward scattering, thus this small difference is inessential and we will approximate $v_{F0}=v_{F\pm1}=v_{F}$ at first.

The bosonized Hamiltonian $H^B$ also consists of two parts,
\begin{equation}\label{Eq:Hamiltonian_B}
H^{B}=H_{0}^{B}+H_{int}^{B}.
\end{equation}
$H_{0}^{B}$ is the quadratic or non-interacting part, and $H_{int}^B$ is the interacting part.
The non-interacting part $H_{0}^{B}$ can be diagonalized by Eq.~(\ref{Eq:diagonalize-HB0}), resulting in
\begin{equation}\label{Eq:H0B}
H_{0}^{B}=\frac{1}{2\pi}\int dx\sum_{\mu\nu}
v_{\mu\nu}\left[K_{\mu\nu}\left(\nabla\tilde{\theta}_{\mu\nu}\right)^{2}+\frac{1}{K_{\mu\nu}}\left(\nabla\tilde{\phi}_{\mu\nu}\right)^{2}\right]
\end{equation}
where $\mu=c,s$ and $\nu=0,\pm1$. The renormalized Fermi velocity $v_{\mu\nu}$ and Tomonaga-Luttinger parameters $K_{\mu\nu}$ are given by
\begin{subequations}\label{Eq:v-K}
\begin{align}
\frac{v_{c\left(s\right)\pm1}}{v_{F}} &=  \sqrt{1-\frac{\left[+\left(-\right)g_{4\perp}^{(2)}-\left(g_{2\parallel}^{(2)}+\left(-\right)g_{2\perp}^{(2)}\right)\right]^{2}}{\left(2\pi v_{F}\right)^{2}}},  \\
\frac{v_{c\left(s\right)0}}{v_{F}} &= \sqrt{ 1-\frac{\left[+\left(-\right)g_{4\perp}^{(2)}+2\left(g_{2\parallel}^{(2)}+\left(-\right)g_{2\perp}^{(2)}\right)\right]^{2}}{\left(2\pi v_{F}\right)^{2}}}, \\
K_{c\left(s\right)\pm1} &= \sqrt{ \frac{1-\frac{1}{2\pi v_{F}}\left[+\left(-\right)g_{4\perp}^{(2)}-\left(g_{2\parallel}^{(2)}+\left(-\right)g_{2\perp}^{(2)}\right)\right]}{1+\frac{1}{2\pi v_{F}}\left[+\left(-\right)g_{4\perp}^{(2)}-\left(g_{2\parallel}^{(2)}+\left(-\right)g_{2\perp}^{(2)}\right)\right]}},  \\
K_{c\left(s\right)0} &= \sqrt{ \frac{1-\frac{1}{2\pi v_{F}}\left[+\left(-\right)g_{4\perp}^{(2)}+2\left(g_{2\parallel}^{(2)}+\left(-\right)g_{2\perp}^{(2)}\right)\right]}{1+\frac{1}{2\pi v_{F}}\left[+\left(-\right)g_{4\perp}^{(2)}+2\left(g_{2\parallel}^{(2)}+\left(-\right)g_{2\perp}^{(2)}\right)\right]}}.
\end{align}
\end{subequations}
The spin-charge separation is reflected in $v_{cm}\neq v_{sm}$, which is similar to single band Tomonaga-Luttinger liquids.
The difference between single-band and three-band model is the following. For the single-band model, all the forward scattering processes contribute to the bosonic non-interacting Hamiltonian $H_0^B$.
However, for the three-band model, some $g^{(2)}$ forward scattering processes contribute to the interacting part $H_{int}^B$ but not the non-interacting part $H_0^B$, which can be seen in Eq.~(\ref{Eq:HB-int}) below.
This difference is due to the fact that there is only partial forward scattering processes that can be expressed in the form of density-density or current-current interaction, renormalizing the Tomonaga-Luttinger parameters $K_{\mu\nu}$.

Note that we have omitted the coupling constants with zero initial values in the derivation of $H_0^B$. These coupling constants do not flow under the RG transformation.
We have also dropped all the $g^{(3)}$ umklapp scattering processes, which is negligible when the fermion system is away from half-filling.
All the $g^{(4)}$ and $f^{(4)}$ scattering processes happen within the same chirality and have small momentum transfer. These small-momentum-transfer terms is irrelevant in the sense of RG.
In fact, the $g^{(4)}$ terms will renormalize both $v_{\mu\nu}$ and $K_{\mu\nu}$. However, if one expands $v_{\mu\nu}$ and $K_{\mu\nu}$ in power of $g^{(4)}$, the first order terms will vanish.
So that we can safely neglect $g^{(4)}$ and $f^{(4)}$ in both $H_0^B$ and $H_{int}^B$ in perturbation RG, which will not change the conclusions of our perturbation RG analysis in remaining parts of this paper.

The bosonic interacting Hamiltonian $H_{int}^B$ is given by
\begin{widetext}
\begin{eqnarray}\label{Eq:HB-int}
H_{int}^{B}= & - & g_{1\perp}^{(1)}\frac{4}{\left(2\pi a\right)^{2}}\int dx\cos\left(\frac{2}{\sqrt{3}}\tilde{\phi}_{s-1}+\frac{4}{\sqrt{6}}\tilde{\phi}_{s0}\right)\cos\left(2\tilde{\theta}_{s+1}\right)\nonumber \\
 & + & g_{2\parallel}^{(1)}\frac{4}{\left(2\pi a\right)^{2}}\int dx\cos\left(2\tilde{\phi}_{c+1}\right)\cos\left(2\tilde{\phi}_{s+1}\right)\nonumber \\
 & + & g_{2\perp}^{(1)}\frac{4}{\left(2\pi a\right)^{2}}\int dx\cos\left(2\tilde{\phi}_{c+1}\right)\cos\left(\frac{2}{\sqrt{3}}\tilde{\phi}_{s-1}+\frac{4}{\sqrt{6}}\tilde{\phi}_{s0}\right)\nonumber \\
 & + & g_{3\parallel}^{(1)}\frac{4}{\left(2\pi a\right)^{2}}\int dx\cos\left(2\tilde{\theta}_{c+1}\right)\cos\left(2\tilde{\theta}_{s+1}\right)\nonumber \\
 & + & g_{3\perp}^{(1)}\frac{4}{\left(2\pi a\right)^{2}}\int dx\cos\left(2\tilde{\theta}_{c+1}\right)\cos\left(\frac{2}{\sqrt{3}}\tilde{\phi}_{s-1}+\frac{4}{\sqrt{6}}\tilde{\phi}_{s0}\right)\nonumber \\
 & + & g_{4\perp}^{(1)}\frac{4}{\left(2\pi a\right)^{2}}\int dx\cos\left(2\tilde{\phi}_{s+1}\right)\cos\left(\frac{2}{\sqrt{3}}\tilde{\phi}_{s-1}+\frac{4}{\sqrt{6}}\tilde{\phi}_{s0}\right)\nonumber \\
 & - & g_{1\perp}^{(2)}\frac{4}{\left(2\pi a\right)^{2}}\int dx\cos\left(2\tilde{\phi}_{c+1}\right)\cos\left(2\tilde{\theta}_{s+1}\right)\nonumber \\
 & + & g_{3\perp}^{(2)}\frac{4}{\left(2\pi a\right)^{2}}\int dx\cos\left(2\tilde{\theta}_{c+1}\right)\cos\left(2\tilde{\phi}_{s+1}\right)\nonumber \\
 & - & f_{1\perp}^{(1)}\frac{8}{\left(2\pi a\right)^{2}}\int dx\left[\cos\tilde{\phi}_{s+1}\cos\left(-\frac{1}{\sqrt{3}}\tilde{\phi}_{s-1}+\frac{4}{\sqrt{6}}\tilde{\phi}_{s0}\right)\cos\tilde{\theta}_{s+1}\cos\sqrt{3}\tilde{\theta}_{s-1}+\left(\cos\rightarrow\sin\right)\right]\nonumber \\
 & + & f_{3\parallel}^{(1)}\frac{8}{\left(2\pi a\right)^{2}}\int dx\left[\cos\tilde{\theta}_{c+1}\cos\sqrt{3}\tilde{\theta}_{c-1}\cos\tilde{\theta}_{s+1}\cos\sqrt{3}\tilde{\theta}_{s-1}+\left(\cos\rightarrow\sin\right)\right]\nonumber \\
 & + & f_{3\perp}^{(1)}\frac{8}{\left(2\pi a\right)^{2}}\int dx\left[\cos\tilde{\theta}_{c+1}\cos\sqrt{3}\tilde{\theta}_{c-1}\cos\tilde{\phi}_{s+1}\cos\left(-\frac{1}{\sqrt{3}}\tilde{\phi}_{s-1}+\frac{4}{\sqrt{6}}\tilde{\phi}_{s0}\right)+\left(\cos\rightarrow\sin\right)\right]\nonumber \\
 & + & f_{3\perp}^{(2)}\frac{8}{\left(2\pi a\right)^{2}}\int dx\left[\cos\tilde{\theta}_{c+1}\cos\sqrt{3}\tilde{\theta}_{c-1}\cos\tilde{\phi}_{s+1}\cos\sqrt{3}\tilde{\phi}_{s-1}+\left(\cos\rightarrow\sin\right)\right]\nonumber \\
 & + & g_{\perp}^{(1)}\frac{2}{\left(2\pi a\right)^{2}}\int dx\cos\left(-\frac{4}{\sqrt{3}}\tilde{\phi}_{s-1}+\frac{4}{\sqrt{6}}\tilde{\phi}_{s0}\right).
\end{eqnarray}
\end{widetext}
Unlike the non-interacting situation in $H_0^B$, here we retain the terms with zero initial values of coupling constants in $H_{int}^B$, namely, $g_{3\parallel}^{(1)}$ and $f_{3\parallel}^{(1)}$.
These terms will be automatically generated in the one-loop RG to form the closed algebra of operator product expansion (OPE).
Since different scattering processes may give rise to the same form in the bosonized Hamiltonian, we have incorporated them into a single term, e.g. $g_{3\parallel}^{(1)}$ and $g_{3\parallel}^{(2)}$ terms.
As mentioned before, both the forward and back scattering processes will contribute to $H_{int}^B$ in the three-band case, which is different from the single-band case.

The non-interacting Hamiltonian $H_0^B$ describes a three-band Tomonaga-Luttinger liquid, which is a Gaussian fixed point under RG transformation.
In the remaining sections, we shall treat the interacting part $H_{int}^B$ as a perturbation and perform RG analysis to investigate its relevance.
The (most) relevant terms in $H_{int}^B$ will give the low energy effective field theories. In such effective field theories, the fields $\phi_{\mu\nu}$ and $\theta_{\mu\nu}$
will be locked around some saddle points, say, the extrema of cosine functions in Eq.~(\ref{Eq:HB-int}), which gives rise to some ordered states or relevant instabilities.
To classify such orders or instabilities, we shall introduce order parameters in the next section at first.

\section{order parameter}\label{order}
To characterize different effective field theories in low energies, we shall introduce order parameters in this section.
In general, the order parameter can be defined as fermionic bilinear, or more precisely, long-ranged correlation of bilinear fermionic operators.
By this definition, there are two classes of order parameters in such a three-band system. One is defined in the particle-hole channels,
\begin{subequations}\label{Eq:OP-def}
\begin{equation}
O_{ph}^{ij} = \sum_{mm'\sigma\sigma'}\lambda_{mm'}^{i}\tau_{\sigma\sigma'}^{j}\psi_{Rm\sigma}^{\dagger}\psi_{Lm'\sigma'},
\end{equation}
and the other is defined in particle-particle channels (or their Hermitian conjugates in hole-hole channels),
\begin{equation}
O_{pp}^{ij} = \sum_{mm'\sigma\sigma'}\sigma\lambda_{mm'}^{i}\tau_{\sigma\sigma'}^{j}\psi_{Rm\sigma}^{\dagger}\psi_{Lm'\bar{\sigma'}}^{\dagger}.
\end{equation}
\end{subequations}
Here $\lambda^{i}(i=1,\cdots,8)$ are Gell-Mann matrices, $\tau^{j}(j=1,2,3)$ are Pauli matrices. We have also defined $\lambda^{0}$ and $\sigma^0$ as $3\times 3$ and $2\times 2$ unit matrices respectively.
$\psi_{pm\sigma}(\psi_{pm\sigma}^{\dagger})$ is the electron annihilation (creation) operator with chirality $p$, band $m$ and spin $\sigma$.

Note that we only keep opposite-chirality terms, $\psi_R^{\dagger}\psi_L$ and $\psi_R^{\dagger}\psi_L^{\dagger}$ in Eq.~(\ref{Eq:OP-def}) and ignore equal-chirality terms,
such as $\psi_R^{\dagger}\psi_R$, $\psi_L^{\dagger}\psi_L$, $\psi_R^{\dagger}\psi_R^{\dagger}$, and $\psi_L^{\dagger}\psi_L^{\dagger}$. This is because four-fermion operators in the same chirality,
e.g. $\psi_R^{\dagger}\psi_R^{\dagger}\psi_R^{\dagger}\psi_R$ and  $\psi_R^{\dagger}\psi_R\psi_R^{\dagger}\psi_R$ are all irrelevant in the sense of RG.
Physically, all our familiar ordered states, including charge density wave (CDW), spin density wave (SDW), and superconducting (SC) states, arise from scattering or pairing in opposite chiralities.
Therefore, Eq.~(\ref{Eq:OP-def}) contains all possible physically relevant order parameters constructed by fermionic bilinears.

We shall identify physical ordered states for each order parameters in Eq.~(\ref{Eq:OP-def}) and bosonize them.
For particle-hole channels, we find that $O_{ph}^{i0}$ refers to CDW and $O_{ph}^{i1-3}$ refer to the three components of SDW.
There are total $9\times 4=36$ order parameters in particle-hole channels. Below we only list 4 $\lambda^1$-components as examples, which involve only two $E^{\prime}$ bands with $m=\pm 1$.
After bosonization, these four order parameters read
\begin{widetext}
\begin{subequations}\label{Eq:OP-ph}
\begin{eqnarray}
O_{ph}^{10} & \propto & e^{-i2k_{F}x+i\left(\frac{1}{\sqrt{3}}\tilde{\phi}_{c-1}+\frac{2}{\sqrt{6}}\tilde{\phi}_{c0}\right)}\left[\cos\left(\frac{1}{\sqrt{3}}\tilde{\phi}_{s-1}+\frac{2}{\sqrt{6}}\tilde{\phi}_{s0}\right)\cos\tilde{\theta}_{c+1}\sin\tilde{\theta}_{s+1}+i\left(\cos\leftrightarrow\sin\right)\right],\\
O_{ph}^{11} & \propto & e^{-i2k_{F}x+i\left(\frac{1}{\sqrt{3}}\tilde{\phi}_{c-1}+\frac{2}{\sqrt{6}}\tilde{\phi}_{c0}\right)}\left[\cos\left(\frac{1}{\sqrt{3}}\tilde{\theta}_{s-1}+\frac{2}{\sqrt{6}}\tilde{\theta}_{s0}\right)\sin\tilde{\theta}_{c+1}\cos\tilde{\phi}_{s+1}+i\left(\cos\leftrightarrow\sin\right)\right],\\
O_{ph}^{12} & \propto & e^{-i2k_{F}x+i\left(\frac{1}{\sqrt{3}}\tilde{\phi}_{c-1}+\frac{2}{\sqrt{6}}\tilde{\phi}_{c0}\right)}\left[\sin\left(\frac{1}{\sqrt{3}}\tilde{\theta}_{s-1}+\frac{2}{\sqrt{6}}\tilde{\theta}_{s0}\right)\sin\tilde{\theta}_{c+1}\cos\tilde{\phi}_{s+1}-i\left(\cos\leftrightarrow\sin\right)\right],\\
O_{ph}^{13} & \propto & e^{-i2k_{F}x+i\left(\frac{1}{\sqrt{3}}\tilde{\phi}_{c-1}+\frac{2}{\sqrt{6}}\tilde{\phi}_{c0}\right)}\left[\cos\left(\frac{1}{\sqrt{3}}\tilde{\phi}_{s-1}+\frac{2}{\sqrt{6}}\tilde{\phi}_{s0}\right)\sin\tilde{\theta}_{c+1}\cos\tilde{\theta}_{s+1}+i\left(\cos\leftrightarrow\sin\right)\right],
\end{eqnarray}
\end{subequations}
\end{widetext}
where $2k_{F}=k_{F+1}+k_{F-1}$, and $(\cos \leftrightarrow \sin)$ means replacing all the cosine functions by sine functions and vice versa.

For particle-particle channels, we find that $O_{pp}^{i0}$ serves as singlet superconducting (SSC) pairing order parameter and $O_{pp}^{i1-3}$ serve as three components of triplet superconducting (TSC) pairing order parameters.
The bosonization for $\lambda^2$-components is the following,
\begin{widetext}
\begin{subequations}\label{Eq:OP-pp}
\begin{eqnarray}
O_{pp}^{20} & \propto & e^{i\left(\frac{1}{\sqrt{3}}\tilde{\theta}_{c-1}+\frac{2}{\sqrt{6}}\tilde{\theta}_{c0}\right)}\left[\cos\left(\frac{1}{\sqrt{3}}\tilde{\phi}_{s-1}+\frac{2}{\sqrt{6}}\tilde{\phi}_{s0}\right)\sin\tilde{\phi}_{c+1}\sin\tilde{\theta}_{s+1}-i\left(\cos\leftrightarrow\sin\right)\right],\\
O_{pp}^{21} & \propto & e^{i\left(\frac{1}{\sqrt{3}}\tilde{\theta}_{c-1}+\frac{2}{\sqrt{6}}\tilde{\theta}_{c0}\right)}\left[\cos\left(\frac{1}{\sqrt{3}}\tilde{\theta}_{s-1}+\frac{2}{\sqrt{6}}\tilde{\theta}_{s0}\right)\cos\tilde{\phi}_{c+1}\cos\tilde{\phi}_{s+1}-i\left(\cos\leftrightarrow\sin\right)\right],\\
O_{pp}^{22} & \propto & e^{i\left(\frac{1}{\sqrt{3}}\tilde{\theta}_{c-1}+\frac{2}{\sqrt{6}}\tilde{\theta}_{c0}\right)}\left[\sin\left(\frac{1}{\sqrt{3}}\tilde{\theta}_{s-1}+\frac{2}{\sqrt{6}}\tilde{\theta}_{s0}\right)\cos\tilde{\phi}_{c+1}\cos\tilde{\phi}_{s+1}+i\left(\cos\leftrightarrow\sin\right)\right],\\
O_{pp}^{23} & \propto & e^{i\left(\frac{1}{\sqrt{3}}\tilde{\theta}_{c-1}+\frac{2}{\sqrt{6}}\tilde{\theta}_{c0}\right)}\left[\cos\left(\frac{1}{\sqrt{3}}\tilde{\phi}_{s-1}+\frac{2}{\sqrt{6}}\tilde{\phi}_{s0}\right)\cos\tilde{\phi}_{c+1}\cos\tilde{\theta}_{s+1}-i\left(\cos\leftrightarrow\sin\right)\right].
\end{eqnarray}
\end{subequations}
\end{widetext}

Each bosonized order parameter contains two parts, which are related to each other by interchanging $\cos\leftrightarrow\sin$.
If one shifts the bosonic fields $\theta_{\mu+1}$ and $\phi_{\mu+1}$ by $\frac{\pi}{\sqrt{2}}$,
\begin{equation}\label{Eq:shift-1}
\phi(\theta)_{\mu+1}  \to  \phi(\theta)_{\mu+1} + \frac{\pi}{\sqrt{2}},
\end{equation}
 and leaves other $\theta_{\mu\nu}$'s and $\phi_{\mu\nu}$'s unchanged, the diagonalized fields $\tilde\theta_{\mu\nu}$ and $\tilde\phi_{\mu\nu}$ will transfer accordingly,
\begin{eqnarray}
\tilde{\phi}(\tilde{\theta})_{\mu+1} & \to & \tilde{\phi}(\tilde{\theta})_{\mu+1} + \frac{\pi}{2}, \nonumber\\
\tilde{\phi}(\tilde{\theta})_{\mu-1} & \to & \tilde{\phi}(\tilde{\theta})_{\mu-1} + \frac{\pi}{2\sqrt{3}}, \nonumber\\
\tilde{\phi}(\tilde{\theta})_{\mu0} & \to & \tilde{\phi}(\tilde{\theta})_{\mu0}+ \frac{\pi}{\sqrt{6}}. \nonumber
\end{eqnarray}
So that
\begin{equation*}
\frac{1}{\sqrt{3}}\tilde{\phi}(\tilde{\theta})_{\mu-1}+\frac{2}{\sqrt{6}}\tilde{\phi}(\tilde{\theta})_{\mu0}  \to  \frac{1}{\sqrt{3}}\tilde{\phi}(\tilde{\theta})_{\mu-1}+\frac{2}{\sqrt{6}}\tilde{\phi}(\tilde{\theta})_{\mu0}+{\pi\over 2}.
\end{equation*}
It means that the order parameters $O_{ph}^{i1-3}$ and $O_{pp}^{i1-3}$  will not change under the phase shift given in Eq.~(\ref{Eq:shift-1}).
This can be verified by the bosonization formula Eq.~(\ref{Eq:bosonization}) too.

In the following RG analysis, the coupling constants will flow to zero if they are irrelevant and to strong coupling limit if they are relevant.
The relevant coupling constants will lock the corresponding bosonic fields $\theta_{\mu\nu}$ and $\phi_{\mu\nu}$ around the saddle points, say, in the extremum of cosine or sine functions to minimize the action.
If we substitute these saddle-point values of bosonic fields into the order parameters, we will obtain the nonzero order parameters. For instance, the saddle point
\begin{equation}
\left(\frac{1}{\sqrt{3}}\tilde{\phi}_{s-1}+\frac{2}{\sqrt{6}}\tilde{\phi}_{s0},\,\tilde{\phi}_{c+1},\,\tilde{\theta}_{s+1}\right)=\left(0,\,0,\,0\right)
\end{equation}
will give rise to nonzero amplitude for order parameter $O_{pp}^{23}$ in Eq.~(\ref{Eq:OP-pp}). The remaining phase factor
$$
e^{i\left(\frac{1}{\sqrt{3}}\tilde{\theta}_{c-1}+\frac{2}{\sqrt{6}}\tilde{\theta}_{c0}\right)}=e^{\frac{i}{\sqrt{2}}\left(\theta_{c+1}+\theta_{c-1}\right)}
$$
reflects the $U(1)$ gauge symmetry, which will be spontaneously broken when the SC long ranged order is established.

\section{Renormalization-Group analysis}\label{RG}

The quadratic part of the Hamiltonian, $H_0^B$, is a well defined Gaussian fixed point under RG, describing the three-band TLLs at high temperatures, which servers as a good starting point for our study.
In this section, we begin with the quadratic (non-interacting) part $H_0^B$ and treat the nonquadratic (interacting) part $H_{int}^B$ by a RG method perturbatively.
We shall use OPE method \cite{JCardy_OPE} to derive the RG equations for the 13 coupling constants in Eq.~(\ref{Eq:HB-int}) up to one loop.

The general form of one-loop perturbative RG equations read
\begin{equation}\label{Eq:general-RGE}
\frac{dg_{k}}{dl}=\left(d-\Delta_{k}\right)g_{k}-\sum_{ij}C_{ij}^{k}g_{i}g_{j},
\end{equation}
where $g_k$ represents the coupling constants ($g$'s and $f$'s) in $H_{int}^B$ in Eq.~(\ref{Eq:HB-int}). The linear term in Eq.~(\ref{Eq:general-RGE}) is the tree-level contribution
and depends on space-time dimension $d$ and scaling dimension $\Delta_{k}$. The quadratic terms are the one-loop contributions. The coefficients $C_{ij}^{k}$ are the structure constants of the OPE,
which can be obtained by the fusion of arbitrary two terms in $H_{int}^B$. This process will generate new terms, which are absent in the original microscopic Hamiltonian, until all terms form a closed algebra.
This is the reason why we retain the terms with zero initial values of coupling constants in Eq.~(\ref{Eq:HB-int}).

In the spirit of perturbation theory, we shall firstly derive and analyze RG equations at tree-level, and then carry out one-loop analysis in the remaining parts of this section.

\subsection{Tree-level RG}\label{sec:tree-RG}

To simplify, we introduce the dimensionless coupling constants
\begin{subequations}\label{Eq:xyi}
\begin{align}
y_{i}&=\frac{g_{i}}{\pi v_{F}}, \\
x_{i}&=\frac{f_{i}}{\pi v_{F}}.
\end{align}
\end{subequations}
As shown in Appendix \ref{App:tree-RG},
the tree-level RG equations in weak coupling can be written in terms of $x_i$ and $y_i$'s,
\begin{subequations}\label{Eq:tree-RG2}
\begin{align}
\frac{dy_{1\perp}^{(1)}}{dl}&=\left(y_{2\parallel}^{(2)}-y_{2\perp}^{(2)}\right)y_{1\perp}^{(1)},
\end{align}
\begin{align}
\frac{dy_{2\parallel}^{(1)}}{dl}&=-y_{2\parallel}^{(2)}y_{2\parallel}^{(1)},
\end{align}
\begin{align}
\frac{dy_{2\perp}^{(1)}}{dl}&=-y_{2\perp}^{(2)}y_{2\perp}^{(1)},
\end{align}
\begin{align}
\frac{dy_{3\parallel}^{(1)}}{dl}&=y_{2\parallel}^{(2)}y_{3\parallel}^{(1)},
\end{align}
\begin{align}
\frac{dy_{3\perp}^{(1)}}{dl}&=\left(-y_{4\perp}^{(2)}+y_{2\parallel}^{(2)}\right)y_{3\perp}^{(1)},
\end{align}
\begin{align}
\frac{dy_{4\perp}^{(1)}}{dl}&=-y_{2\perp}^{(2)}y_{4\perp}^{(1)},
\end{align}
\begin{align}
\frac{dy_{1\perp}^{(2)}}{dl}&=\left(y_{4\perp}^{(2)}-y_{2\perp}^{(2)}\right)y_{1\perp}^{(2)},
\end{align}
\begin{align}
\frac{dy_{3\perp}^{(2)}}{dl}&=\left(-y_{4\perp}^{(2)}+y_{2\perp}^{(2)}\right)y_{3\perp}^{(2)},
\end{align}
\begin{align}
\frac{dx_{1\perp}^{(1)}}{dl}&=\left(y_{2\parallel}^{(2)}-y_{2\perp}^{(2)}\right)x_{1\perp}^{(1)},
\end{align}
\begin{align}
\frac{dx_{3\parallel}^{(1)}}{dl}&=y_{2\parallel}^{(2)}x_{3\parallel}^{(1)},
\end{align}
\begin{align}
\frac{dx_{3\perp}^{(1)}}{dl}&=\left(-y_{4\perp}^{(2)}+y_{2\parallel}^{(2)}\right)x_{3\perp}^{(1)},
\end{align}
\begin{align}
\frac{dx_{3\perp}^{(2)}}{dl}&=\left(-y_{4\perp}^{(2)}+y_{2\perp}^{(2)}\right)x_{3\perp}^{(2)},
\end{align}
\begin{align}
\frac{dy_{\perp}^{(1)}}{dl}&=-y_{4\perp}^{(2)}y_{\perp}^{(1)}.
\end{align}
\end{subequations}

In the formulation of Abelian bosonization, Eqs.~(\ref{Eq:bosonization}), the variables $y_{2\parallel}^{(2)}$, $y_{2\perp}^{(2)}$ and $y_{4\perp}^{(2)}$ in Eqs.~(\ref{Eq:K-y}) and (\ref{Eq:tree-RG2})
only appear in the quardratic part $H_0^B$ in the original microscopic Hamiltonian, thus do not flow under the RG transformation. So that we use their initial values
$y_{2\parallel}^{(2)}=\frac{U-3J}{\pi v_F}$, $y_{2\perp}^{(2)}=\frac{U-2J}{\pi v_F}$ and $y_{4\perp}^{(2)}=\frac{U}{\pi v_F}$ in tree-level analysis.
However, this formulation does not conserve spin rotational symmetry. We shall discuss how to restore spin $SU(2)$ symmetry in the next subsection, where $y_{2\parallel}^{(2)}$, $y_{2\perp}^{(2)}$ and $y_{4\perp}^{(2)}$
can be expressed in terms of the 13 coupling constants in Eq.~(\ref{Eq:HB-int}) and make the RG equations close.

The slope $\{\frac{1}{x_i}\frac{d x_i}{dl},\frac{1}{y_i}\frac{d y_i}{dl}\}$ around the Tomonaga-Luttinger liquid fixed point will determine which coupling constants are relevant.
In weak coupling, this slope is given by the initial values of the coupling constants in Eq.~(\ref{Eq:couplings}), say, the microscopic Hamiltonian with two parameters $U>0$ and $J>0$. We find that there exist three parameters regions.
(1) For $0<J<U/3$, the coupling constants $x_{3\parallel}^{(1)}$, $y_{3\parallel}^{(1)}$ and $y_{1\perp}^{(2)}$ are relevant,
other coupling constants are irrelevant.
(2) For $U/3<J<U/2$, there are only two relevant coupling constants, $y_{2\parallel}^{(1)}$ and $y_{1\perp}^{(2)}$.
(3) For the unphysical region $J>U/2$, there are four relevant coupling constants, $y_{2\parallel}^{(1)}$, $y_{2\perp}^{(1)}$, $y_{4\perp}^{(1)}$ and $y_{1\perp}^{(2)}$.
However, the above analysis  relies largely on tree-level RG equations. We now proceed to one-loop RG equations for further study.

\subsection{One-loop RG}\label{sec:oneloop-RG}

With the help of spin $SU(2)$ symmetry and the microscopic Hamiltonian, we are able to derive one-loop RG equations (see Appendix~\ref{App:oneloop-RG}) as follows,
\begin{subequations}\label{Eq:one-loopRGE}
\begin{align}\label{Eq:y1perp1}
\frac{dy_{1\perp}^{(1)}}{dl}  =  -\left(y_{1\perp}^{(1)}\right)^{2}-y_{2\perp}^{(1)}y_{1\perp}^{(2)}+y_{3\parallel}^{(1)}y_{3\perp}^{(1)},
\end{align}
\begin{align}\label{Eq:y2para1}
\frac{dy_{2\parallel}^{(1)}}{dl}  =  \frac{1}{2}y_{1\perp}^{(1)}y_{2\parallel}^{(1)}-y_{2\perp}^{(1)}y_{4\perp}^{(1)},
\end{align}
\begin{align}\label{Eq:y2perp1}
\frac{dy_{2\perp}^{(1)}}{dl}  =  -\frac{1}{2}y_{1\perp}^{(1)}y_{2\perp}^{(1)}-y_{1\perp}^{(1)}y_{1\perp}^{(2)} -y_{2\parallel}^{(1)}y_{4\perp}^{(1)},
\end{align}
\begin{align}\label{Eq:y3para1}
\frac{dy_{3\parallel}^{(1)}}{dl}  =  -\frac{1}{2}y_{1\perp}^{(1)}y_{3\parallel}^{(1)}+y_{1\perp}^{(1)}y_{3\perp}^{(1)},
\end{align}
\begin{align}\label{Eq:y3perp1}
\frac{dy_{3\perp}^{(1)}}{dl}  = & -\left(y_{4\perp}^{(1)}+\frac{1}{2}y_{1\perp}^{(1)}\right)y_{3\perp}^{(1)} +y_{1\perp}^{(1)}y_{3\parallel}^{(1)}\nonumber\\
                                & -y_{4\perp}^{(1)}y_{3\perp}^{(2)},
\end{align}
\begin{align}\label{Eq:y4perp1}
\frac{dy_{4\perp}^{(1)}}{dl}  =  \frac{1}{2}y_{1\perp}^{(1)}y_{4\perp}^{(1)}-y_{2\parallel}^{(1)}y_{2\perp}^{(1)} -y_{3\perp}^{(1)}y_{3\perp}^{(2)},
\end{align}
\begin{align}\label{Eq:y1perp2}
\frac{dy_{1\perp}^{(2)}}{dl}  =  \left(y_{4\perp}^{(1)}-\frac{1}{2}y_{1\perp}^{(1)}\right)y_{1\perp}^{(2)} -y_{1\perp}^{(1)}y_{2\perp}^{(1)},
\end{align}
\begin{align}\label{Eq:y3perp2}
\frac{dy_{3\perp}^{(2)}}{dl}  =  \left(-y_{4\perp}^{(1)}+\frac{1}{2}y_{1\perp}^{(1)}\right)y_{3\perp}^{(2)} -y_{3\perp}^{(1)}y_{4\perp}^{(1)},
\end{align}
\begin{align}\label{Eq:x1perp1}
\frac{dx_{1\perp}^{(1)}}{dl}  =  -\left(x_{1\perp}^{(1)}\right)^{2}+x_{3\parallel}^{(1)}x_{3\perp}^{(1)},
\end{align}
\begin{align}\label{Eq:x3para1}
\frac{dx_{3\parallel}^{(1)}}{dl}  =  -\frac{1}{2}x_{1\perp}^{(1)}x_{3\parallel}^{(1)}+x_{1\perp}^{(1)}x_{3\perp}^{(1)},
\end{align}
\begin{align}\label{Eq:x3perp1}
\frac{dx_{3\perp}^{(1)}}{dl}  =  -\left(y_{\perp}^{(1)}+\frac{1}{2}x_{1\perp}^{(1)}\right)x_{3\perp}^{(1)} +x_{1\perp}^{(1)}x_{3\parallel}^{(1)}-y_{\perp}^{\left(1\right)}x_{3\perp}^{\left(2\right)},
\end{align}
\begin{align}\label{Eq:x3perp2}
\frac{dx_{3\perp}^{(2)}}{dl}  =  \left(-y_{\perp}^{(1)}+\frac{1}{2}x_{1\perp}^{(1)}\right)x_{3\perp}^{(2)}-y_{\perp}^{\left(1\right)}x_{3\perp}^{\left(1\right)},
\end{align}
\begin{align}\label{Eq:y1perp}
\frac{dy_{\perp}^{(1)}}{dl}  =  -\left(y_{\perp}^{(1)}\right)^{2}-x_{3\perp}^{\left(1\right)}x_{3\perp}^{\left(2\right)}.
\end{align}
\end{subequations}
The above 13 RG equations can be classified into two categories. The first eight equations, Eq.~(\ref{Eq:y1perp1}) to Eq.~(\ref{Eq:y3perp2}), describe the RG flow of coupling constants within the two degenerate $E^{\prime}$ bands,
which coincide with those derived in the two-leg-ladder model\cite{EOrignac_twochain}.
The last five equations, Eq.~(\ref{Eq:x1perp1}) to Eq.~(\ref{Eq:y1perp}), couple the two $E^{\prime}$ bands to the non-degenerate $A_{1}^{\prime}$ band.
Note that the last five RG equations are decoupled from the first eight ones. This will greatly simplify our analysis. Such decoupling originates from the particular form of the Hamiltonian \eqref{Eq:Hubbard}, which satisfies
Eq.~\eqref{constraint_added}.

The key to analyze these one-loop RG equations is to find fixed points, where the coupling constants will no longer flow under RG transformation \cite{AAltland_RG}. We rewrite the RG equations in vector form
\begin{equation}
\frac{d\vec{y}}{dl}\equiv \vec{R}\left(\vec{y}\right),
\end{equation}
where $\vec{y}=\left\{ y_{i}\right\}$ is the vector of 13 running coupling constants, and $\vec{R}\left(\vec{y}\right)$ is a vector function of $\vec{y}$. By definition, the fixed points $\vec{y}=\vec{y}^{*}$ are given by
\begin{equation}\label{Eq:fixed point}
\vec{R}\left(\vec{y}^{*}\right)=0.
\end{equation}
It is obvious that $\vec{y}^{*}=0$ is the trivial Tomonaga-Luttinger liquid fixed point. Nontrivial fixed points $\vec{y}^{*}\neq 0$ can be found in perturbation as follows.

In perturbation, we are able to find nontrivial fixed points in two different parameter regions of the microscopic Hamiltonian.

(1) For $0<J<U/3$, we have nontrivial fixed points characterized by the following nonvanishing coupling constants,
\begin{eqnarray}
y_{3\parallel}^{(1)} & = & y_{3\parallel}^{(1)*},\nonumber \\
y_{1\perp}^{(2)} & = & y_{1\perp}^{(2)*}\nonumber, \\
x_{3\parallel}^{(1)} & = & x_{3\parallel}^{(1)*},
\end{eqnarray}
while other coupling constants are all zero.

(2) For $J>U/3$, nontrivial fixed points are given by
\begin{eqnarray}
y_{2\parallel}^{(1)} & = & y_{2\parallel}^{(1)*},\nonumber \\
y_{1\perp}^{(2)} & = & y_{1\perp}^{(2)*},
\end{eqnarray}
while other coupling constants equal zero.

These nontrivial fixed points form hypersurfaces in the 13-dimensional parameter space of coupling constants. By examining the RG flow around these hypersurfaces, we find that these fixed points are phase transition points
rather than stable fixed points describing stable phases. Then we shall analyze the RG flow near the fixed points using one-loop RG equations to find out what kind of instabilities are favored.

In the vicinity of the fixed points, the RG equations can be expanded to linear order
\begin{equation}
\vec{R}\left(\vec{y}\right)=\vec{R}\left(\left(\vec{y}-\vec{y}^{*}\right)+\vec{y}^{*}\right)\simeq W\left(\vec{y}-\vec{y}^{*}\right),
\end{equation}
where the $W$ matrix is defined as
\begin{equation}\label{eq:defW}
W_{ab}=\frac{\partial R_{a}}{\partial y_{b}}|_{\vec{y}=\vec{y}^{*}}.
\end{equation}
We diagonalize the $W$ matrix with the left-eigenvectors $\phi_{\alpha}$,
\begin{equation}
\phi_{\alpha}^{T}W=\phi_{\alpha}^{T}\lambda_{a},
\end{equation}
where $\lambda_{\alpha}$ are corresponding eigenvalues. The scaling fields are defined as
\begin{equation}
v_{\alpha}=\phi_{\alpha}^{T}\left(\vec{y}-\vec{y}^{*}\right).
\end{equation}
Under RG these scaling fields show different behaviors,
\begin{eqnarray}
\frac{dv_{\alpha}}{dl} & = &\phi_{\alpha}^{T}\frac{d}{dl}\left(\vec{y}-\vec{y}^{*}\right)=\phi_{\alpha}^{T}W\left(\vec{y}-\vec{y}^{*}\right) = \lambda_{a}\phi_{\alpha}^{T}\left(\vec{y}-\vec{y}^{*}\right) \nonumber\\
                       & = &\lambda_{a}v_{\alpha},
\end{eqnarray}
which becomes relevant, irrelevant and marginal when $\lambda_{\alpha}>0$,  $\lambda_{\alpha}<0$ and  $\lambda_{\alpha}=0$, respectively.

Note that $\{y_{1\perp}^{(1)}, y_{2\parallel}^{(1)}, y_{2\perp}^{(1)}, y_{3\parallel}^{(1)}, y_{3\perp}^{(1)}, y_{4\perp}^{(1)}, y_{1\perp}^{(2)}, y_{3\perp}^{(2)}\}$
and $\{x_{1\perp}^{(1)}, x_{3\parallel}^{(1)}, x_{3\perp}^{(1)},x_{3\perp}^{(2)},y_{\perp}^{(1)} \}$ form two separated sets in Eqs.~(\ref{Eq:one-loopRGE}). The $W$ matrix is block diagonal as follows,
\begin{equation}\label{Eq:W12}
W=\left(\begin{array}{cc}
W_1 & 0 \\
0 & W_2
\end{array}\right),
\end{equation}
where $W_1$ is a $8\times 8$ matrix and $W_2$ is a $5\times 5$ matrix. The generic forms for $W_1$ and $W_2$ can be found in Appendix \ref{App:dual}.

To illustrate how to carry out the analysis, we firstly consider a simplified case, say, special fixed points when $J>U/3$,
\begin{equation}
y_{1\perp}^{(2)}=y_{1\perp}^{(2)*},
\end{equation}
with other coupling constants equal to zero. In this case, $W_2=0$, and $W_1$ matrix reads
\begin{equation}
W_1=\left(\begin{array}{cccccccc}
0 & 0 & -y_{1\perp}^{(2)*} & 0 & 0 & 0 & 0 & 0 \\
0 & 0 & 0 & 0 & 0 & 0 & 0 & 0 \\
-y_{1\perp}^{(2)*} & 0 & 0 & 0 & 0 & 0 & 0 & 0 \\
0 & 0 & 0 & 0 & 0 & 0 & 0 & 0 \\
0 & 0 & 0 & 0 & 0 & 0 & 0 & 0 \\
0 & 0 & 0 & 0 & 0 & 0 & 0 & 0 \\
-\frac{1}{2}y_{1\perp}^{(2)*} & 0 & 0 & 0 & 0 & y_{1\perp}^{(2)*} & 0 & 0 \\
0 & 0 & 0 & 0 & 0 & 0 & 0 & 0
\end{array}\right).
\end{equation}
For this non-symmetric matrix, we find that there are only two nonzero eigenvalues, $-y_{1\perp}^{(2)*}$ with corresponding eigenvector $y_{1\perp}^{(1)}+y_{2\perp}^{(1)}$
and $y_{1\perp}^{(2)*}$ with eigenvector $y_{1\perp}^{(1)}-y_{2\perp}^{(1)}$.
According to the microscopic model, the initial value $g_{1\perp}^{(2)}=J>0$. Then we expect $y_{1\perp}^{(2)*}>0$ in the RG. Thus the relevant scaling field is given by the eigenvector corresponding to the eigenvalue $y_{1\perp}^{(2)*}$, say,
\begin{equation}
y_{1\perp}^{(1)}-y_{2\perp}^{(1)}.
\end{equation}
Then we can extract relevant terms from the bosonic Hamiltonian $H_{int}^B$. Thus the low energy effective interacting Hamiltonian becomes
\begin{widetext}
\begin{eqnarray}
H_{int}^{B} & = & -\left(g_{1\perp}^{(1)}-g_{2\perp}^{(1)}\right)\frac{2}{\left(2\pi a\right)^{2}}\int dx\cos\left(\frac{2}{\sqrt{3}}\tilde{\phi}_{s-1}+\frac{4}{\sqrt{6}}\tilde{\phi}_{s0}\right)\cos\left(2\tilde{\theta}_{s+1}\right)\nonumber \\
 &  & -\left(g_{1\perp}^{(1)}-g_{2\perp}^{(1)}\right)\frac{2}{\left(2\pi a\right)^{2}}\int dx\cos\left(2\tilde{\phi}_{c+1}\right)\cos\left(\frac{2}{\sqrt{3}}\tilde{\phi}_{s-1}+\frac{4}{\sqrt{6}}\tilde{\phi}_{s0}\right).
\end{eqnarray}
\end{widetext}
When $J>U/3$, the initial value of $y_{1\perp}^{(1)}-y_{2\perp}^{(1)} \propto 3J-U$ is positive. This relevant scaling field will flow to strong coupling and lock the corresponding bosonic fields around the saddle points,
\begin{equation}\label{Eq:saddle-point1}
\begin{array}{ccc}
\left(\frac{1}{\sqrt{3}}\tilde{\phi}_{s-1}+\frac{2}{\sqrt{6}}\tilde{\phi}_{s0},\,\tilde{\phi}_{c+1},\,\tilde{\theta}_{s+1}\right) & = & \left(0,\,0,\,0\right)\\
& \mbox{or} & \left(\frac{\pi}{2},\,\frac{\pi}{2},\,\frac{\pi}{2}\right).
\end{array}
\end{equation}
As discussed following Eq.~(\ref{Eq:shift-1}), these two saddle points will give rise to the same physical states.
The nonzero order parameter corresponding to these locked bosonic fields is $O_{pp}^{23}$, which describes a TSC phase.

Let us turn to generic situations now. We shall neglect vanishing components in $\vec{y}$ for short, and denote $\vec{y}$ as
$\left(y_{3\parallel}^{(1)},y_{1\perp}^{(2)},x_{3\parallel}^{(1)}\right)$ and $\left(y_{1\perp}^{(2)},y_{2\parallel}^{(1)}\right)$ for $0<J<U/3$ and $J>U/3$ respectively.

(1) For $J>U/3$, all the fixed points are in a plane. We can generalize the above analysis for fixed points with two nonzero components, $\left(y_{2\parallel}^{(1)},y_{1\perp}^{(2)}\right)=\left(y_{2\parallel}^{(1)*},y_{1\perp}^{(2)*}\right)$.
In this situation, we still have $W_2=0$, while $W_1$ becomes
\begin{equation}
W_1=\left(\begin{array}{cccccccc}
0 & 0 & -y_{1\perp}^{(2)*} & 0 & 0 & 0 & 0 & 0 \\
\frac{1}{2}y_{2\parallel}^{(1)*} & 0 & 0 & 0 & 0 & 0 & 0 & 0 \\
-y_{1\perp}^{(2)*} & 0 & 0 & 0 & 0 & -y_{2\parallel}^{(1)*} & 0 & 0 \\
0 & 0 & 0 & 0 & 0 & 0 & 0 & 0 \\
0 & 0 & 0 & 0 & 0 & 0 & 0 & 0 \\
0 & 0 & -y_{2\parallel}^{(1)*} & 0 & 0 & 0 & 0 & 0 \\
-\frac{1}{2}y_{1\perp}^{(2)*} & 0 & 0 & 0 & 0 & y_{1\perp}^{(2)*} & 0 & 0 \\
0 & 0 & 0 & 0 & 0 & 0 & 0 & 0
\end{array}\right).
\end{equation}
Now we have four eigenvectors with two corresponding to the zero eigenvalue and the other two corresponding to nonzero eigenvalues. The two nonzero eigenvalues are $\pm \sqrt{\left(y_{2\parallel}^{(1)*}\right)^2+\left(y_{1\perp}^{(2)*}\right)^2}$.
The eigenvector corresponding the positive eigenvalue $\sqrt{\left(y_{2\parallel}^{(1)*}\right)^2+\left(y_{1\perp}^{(2)*}\right)^2}$ is
\begin{align}
y_{1\perp}^{(2)*} y_{1\perp}^{(1)} - \sqrt{\left(y_{2\parallel}^{(1)*}\right)^2+\left(y_{1\perp}^{(2)*}\right)^2} y_{2\perp}^{(1)} + y_{2\parallel}^{(1)*} y_{4\perp}^{(1)}.
\end{align}

Considering the initial value $g_{2\parallel}^{(1)}=U-3J<0$, we expect that $y_{2\parallel}^{(1)*}<0$ for the same reason of perturbation.
In the limit $y_{1\perp}^{(2)*}\to 0$, the relevant scaling field will become $y_{2\perp}^{(1)}+ y_{4\perp}^{(1)}$, which will flow to a strong coupling limit too.
The corresponding saddle point gives rise to a SDW state with order parameter $O_{ph}^{03}+\frac{\sqrt{3}}{2}O_{ph}^{83}$.
In the other limit $y_{2\parallel}^{(1)*}\to 0$, the eigenvector will become $ y_{1\perp}^{(1)}\pm y_{2\perp}^{(1)}$. Then we restore the simplified situation,
where the TSC instability dominates with the order parameter $O_{pp}^{23}$.

\begin{figure}[hptb]
\begin{centering}
\includegraphics[width=8.0cm]{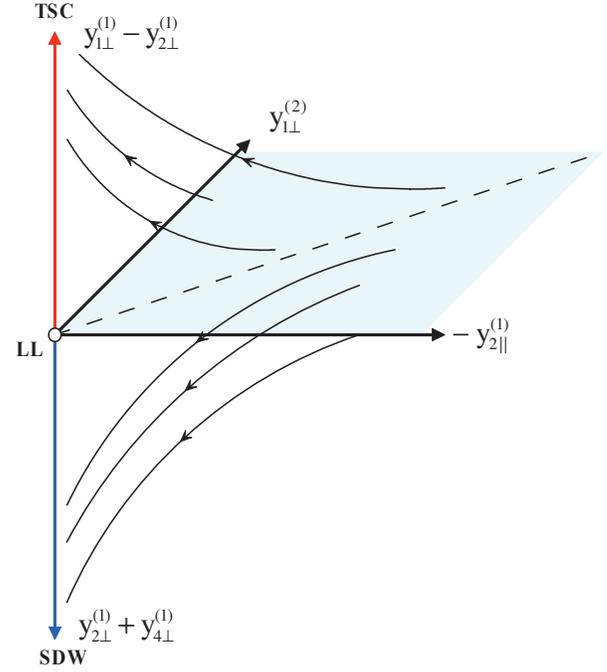}
\caption{(Color online) Sketched RG flow for $J>U/3$. Fixed points ($y_{2\parallel}^{(1)*}<0$ and $y_{1\perp}^{(2)*}>0$) form a quarter plane. The origin is the trivial Tomonaga-Luttinger liquid (TLL) fixed point.
The dashed line denotes the phase boundary separating two phases, TSC and SDW.}
\label{fig:RG_flow}
\end{centering}
\end{figure}

Starting from fixed points between the above two limits, which form a quarter plane $\left(y_{2\parallel}^{(1)*}<0,y_{1\perp}^{(2)*}>0\right)$, the RG trajectory will flow to one of the two strong coupling limits, SDW and TSC.
There must be a phase boundary separating the SDW phase (with order parameter $O_{ph}^{03}+\frac{\sqrt{3}}{2}O_{ph}^{83}$) from the TSC phase (with order parameter $O_{pp}^{23}$).
The RG flow diagram is skectched in Fig.~\ref{fig:RG_flow}. And all the possible ordered ground states for $J>U/3$ are summarized in Table~\ref{table:$U<3J$}.

There exist two competing phases, SDW and TSC, when $J>U/3$. Which one will win out is governed by the microscopic model, say, initial values of the coupling constants.
The most relevant (strongest) instability is given by the largest eigenvalue of $W$ matrix. Assuming that $\vec{y}^{*}$ close to the initial value of $\vec{y}$,
we estimate that the SDW state ($O_{ph}^{03}+\frac{\sqrt{3}}{2}O_{ph}^{83}$) will dominate when $J>U/2$.
And the TSC state ($O_{pp}^{23}$) will become dominant in the region $U/3<J<U/2$.

\begin{table*}[hptb]
\begin{center}
\caption{Possible ordered ground states from one-loop RG analysis when $J>U/3$.}
\label{table:$U<3J$}
\begin{tabular}{|c|c|c|}
\hline
\rule{0pt}{1.5em} Scaling field & $y_{2\perp}^{(1)}+y_{4\perp}^{(1)}$ & $y_{1\perp}^{(1)}-y_{2\perp}^{(1)}$\tabularnewline
\hline
\rule{0pt}{1.5em} Instability & SDW & TSC\tabularnewline
\hline
\rule{0pt}{1.5em} Order parameter & $O_{ph}^{03}+\frac{\sqrt{3}}{2}O_{ph}^{83}$ & $O_{pp}^{23}$\tabularnewline
\hline
\rule{0pt}{3.5em} Saddle point & $\begin{array}{ccc}
\frac{1}{\sqrt{3}}\tilde{\phi}_{s-1}+\frac{2}{\sqrt{6}}\tilde{\phi}_{s0} & = & 0 \left(\frac{\pi}{2}\right)\\
\tilde{\phi}_{c+1} & = & \frac{\pi}{2} \left(0\right)\\
\tilde{\phi}_{s+1} & = & \frac{\pi}{2} \left(0\right)
\end{array}$ & $\begin{array}{ccc}
\frac{1}{\sqrt{3}}\tilde{\phi}_{s-1}+\frac{2}{\sqrt{6}}\tilde{\phi}_{s0} & = & 0 \left(\frac{\pi}{2}\right)\\
\tilde{\phi}_{c+1}   & = & 0 \left(\frac{\pi}{2}\right)\\
\tilde{\theta}_{s+1} & = & 0 \left(\frac{\pi}{2}\right)
\end{array}$\tabularnewline
\hline
\end{tabular}
\end{center}
\end{table*}

(2) For $0<J<U/3$, we have three non-vanishing components $\left(y_{3\parallel}^{(1)},y_{1\perp}^{(2)},x_{3\parallel}^{(1)}\right)$ in $\vec{y}^{*}$, which form a three dimensional hypersurface.
In this case, we have $W_1$ and $W_2$ matrices as follows,
\begin{subequations}
\begin{equation}
W_1=\left(\begin{array}{cccccccc}
0 & 0 & -y_{1\perp}^{(2)*} & 0 & y_{3\parallel}^{(1)*} & 0 & 0 & 0 \\
0 & 0 & 0 & 0 & 0 & 0 & 0 & 0 \\
-y_{1\perp}^{(2)*} & 0 & 0 & 0 & 0 & 0 & 0 & 0 \\
-\frac{1}{2}y_{3\parallel}^{(1)*} & 0 & 0 & 0 & 0 & 0 & 0 & 0 \\
y_{3\parallel}^{(1)*} & 0 & 0 & 0 & 0 & 0 & 0 & 0 \\
0 & 0 & 0 & 0 & 0 & 0 & 0 & 0 \\
-\frac{1}{2}y_{1\perp}^{(2)*} & 0 & 0 & 0 & 0 & y_{1\perp}^{(2)*} & 0 & 0 \\
0 & 0 & 0 & 0 & 0 & 0 & 0 & 0
\end{array}\right),
\end{equation}
and
\begin{equation}
W_2=\left(\begin{array}{ccccc}
0 & 0 & x_{3\parallel}^{(1)*} & 0 & 0 \\
-\frac{1}{2}x_{3\parallel}^{(1)*} & 0 & 0 & 0 & 0 \\
x_{3\parallel}^{(1)*} & 0 & 0 & 0 & 0 \\
0 & 0 & 0 & 0 & 0 \\
0 & 0 & 0 & 0 & 0
\end{array}\right).
\end{equation}
\end{subequations}
These matrices have five nonzero eigenvalues, with three from $W_1$ and two from $W_2$. The three nonzero eigenvalues from $W_1$ are $y_{1\perp}^{(2)*}$, $\pm\sqrt{\left(y_{3\parallel}^{(1)*}\right)^2+\left(y_{1\perp}^{(2)*}\right)^2}$,
and the two from $W_2$ are $\pm x_{3\parallel}^{(1)*}$. Since the related initial values are  $g_{3\parallel}^{(1)}=0$, $g_{1\perp}^{(2)}=J$, and $f_{3\parallel}^{(1)}=0$. We expect $y_{1\perp}^{(2)*}>0$ in perturbation theory.
Moreover, considering the one-loop RG flow around the TLL fixed point, we can deduce that $y_{3\parallel}^{(1)*}>0$ and $x_{3\parallel}^{(1)*}>0$.

Similar analysis can be carried out as in the situation when $J>U/3$. The RG trajectory will flow to three strong coupling limits with relevant scaling fields, $x_{1\perp}^{(1)}+x_{3\perp}^{(1)}$, $y_{1\perp}^{(1)}+y_{3\perp}^{(1)}$, and $y_{1\perp}^{(1)}-y_{2\perp}^{(1)}$.
These relevant scaling fields are associated with positive eigenvalues of $W$ matrix, $x_{3\parallel}^{(1)*}$, $y_{3\parallel}^{(1)*}$ and $y_{1\perp}^{(2)*}$ respectively.
They give rise to two different SDW states (one with order parameter $O_{ph}^{43}$ and $O_{ph}^{63}$ the other with order parameter $O_{ph}^{13}$) and one spin-singlet SC (SSC) state (with order parameter $O_{pp}^{20}$).
All these possible ordered ground states for $0<J<U/3$ and corresponding order parameters are summarized in Table~\ref{table:$U>3J$}.

Then we will compare these three instabilities and find out the strongest one, which is determined by the largest eigenvalue of $W$ matrix.
In the spirit of perturbation theory, we still assume that $\vec{y}^{*}$ close to the initial value of $\vec{y}$.
Note that related initial values are $g_{3\parallel}^{(1)}=f_{3\parallel}^{(1)}=0$ and $g_{1\perp}^{(2)}=J$, we conclude that the SSC state with order parameter $O_{pp}^{20}$ will dominate among these three possible ground states.

\begin{table*}[hptb]
\begin{center}
\caption{Possible ordered ground states from one-loop RG analysis when $0<J<U/3$.}
\label{table:$U>3J$}
\begin{tabular}{|c|c|c|c|}
\hline
\rule{0pt}{1.5em} Scaling field & $x_{1\perp}^{(1)}+x_{3\perp}^{(1)}$ & $y_{1\perp}^{(1)}+y_{3\perp}^{(1)}$ & $y_{1\perp}^{(1)}-y_{2\perp}^{(1)}$\tabularnewline
\hline
\rule{0pt}{1.5em} Instability & SDW & SDW & SSC\tabularnewline
\hline
\rule{0pt}{1.5em} Order parameter & $O_{ph}^{43},O_{ph}^{63}$ & $O_{ph}^{13}$ & $O_{pp}^{20}$\tabularnewline
\hline
\rule{0pt}{3.5em} Saddle point & $\begin{array}{ccc}
\frac{1}{2}\tilde{\phi}_{s+1}-\frac{1}{\sqrt{12}}\tilde{\phi}_{s-1}+\frac{2}{\sqrt{6}}\tilde{\phi}_{s0} & = & 0 \left(\frac{\pi}{2}\right)\\
\frac{1}{2}\tilde{\theta}_{c+1}+\frac{3}{\sqrt{12}}\tilde{\theta}_{c-1} & = & \frac{\pi}{2} \left(0\right)\\
\frac{1}{2}\tilde{\theta}_{s+1}+\frac{3}{\sqrt{12}}\tilde{\theta}_{s-1} & = & 0 \left(\frac{\pi}{2}\right)
\end{array}$ & $\begin{array}{ccc}
\frac{1}{\sqrt{3}}\tilde{\phi}_{s-1}+\frac{2}{\sqrt{6}}\tilde{\phi}_{s0} & = & 0 \left(\frac{\pi}{2}\right)\\
\tilde{\theta}_{c+1} & = & \frac{\pi}{2} \left(0\right)\\
\tilde{\theta}_{s+1} & = & 0 \left(\frac{\pi}{2}\right)
\end{array}$ & $\begin{array}{ccc}
\frac{1}{\sqrt{3}}\tilde{\phi}_{s-1}+\frac{2}{\sqrt{6}}\tilde{\phi}_{s0} & = & 0 \left(\frac{\pi}{2}\right)\\
\tilde{\phi}_{c+1}   & = & \frac{\pi}{2} \left(0\right)\\
\tilde{\theta}_{s+1} & = & \frac{\pi}{2} \left(0\right)
\end{array}$\tabularnewline
\hline
\end{tabular}
\end{center}
\end{table*}

\begin{figure}
\begin{center}
\includegraphics[width=8.4cm]{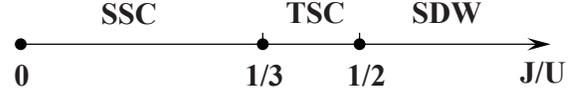}
\caption{Phase diagram for the three-band Hubbard model with two degenerate $E^{\prime}$ orbitals, $k_{F+1}=k_{F-1}$.}
\label{fig:phase_diagram1}
\end{center}
\end{figure}

Let us now summarize the one-loop RG analysis by using OPE and present the phase diagram which are listed in Table~\ref{table:$U<3J$} and \ref{table:$U>3J$}.
In the region $0<J<U/3$, the most relevant instability is the spin-singlet SC instability with order parameter $O_{pp}^{20}$.  At $U/3<J<U/2$, the spin-triplet SC instability with order parameter $O_{pp}^{23}$ is favored.
In the region $J>U/2$ (since $U=U^{\prime}+2J$, $U^{\prime} <0$ in this region), the SDW instability with order parameter $O_{ph}^{03}$ will dominate. The phase diagram is shown in Fig.~\ref{fig:phase_diagram1}.

It worth noting that the one-loop RG Eqs.~\eqref{Eq:one-loopRGE} have been obtained and solved perturbatively in this subsection.
For comparison, we have reproduced established results for two-degenerate-band model\cite{AJMillis_twoband} to examine the validity of this method.
Indeed, there exist other fixed points that is beyond this perturabtive approach. An example for non-perturbative solution is given in Appendix \ref{App:dual}.
We shall also solve the one-loop RG Eqs.~\eqref{Eq:one-loopRGE} numerically in Appendix \ref{App:NRG} to further confirm present results.

\section{Discussions and conclusions}\label{conclusion}

Now we shall relate our theory to experimental results of K$_{2}$Cr$_{3}$As$_{3}$. Firstly, we would like to discuss NMR and NQR experiments.
The spin-lattice relaxation rate $1/T_{1}$ in a NMR experiment measures the local spin correlation which sums over $q$ in the momentum space. The dominant contribution comes from $q\sim 0$ and $q\sim 2k_F$ components.
For a three-band Tomonaga-Luttinger liquid governed by the Hamiltonian $H_0^B$ in Eq.~(\ref{Eq:H0B}), we have the following temperature dependence of $1/T_{1}$ (see Appendix~\ref{App:T1} for details),
\begin{equation}
\frac{1}{T_{1}}\propto A~T + B~T^{1-\frac{U}{2\pi v_F}},
\end{equation}
where $U$ is the effective on-site intra-orbital electron interaction.
The first linearly temperature dependent term follows Korringa law as in Fermi liquids. The second term follows power law with a non-integer exponent as long as $U\neq 0$.
When electron Coulomb repulsion governs the system, $U$ is positive, the dominant contribution at low temperatures will come from the second term.
The spin-lattice relaxation rate $1/T_{1}$ will exhibit non-integer power law temperature dependence.
However, $U$ may become negative effectively, e.g., when electron-phonon interaction dominates over Coulomb repulsion. In this case, $1/T_{1}$ will become linearly temperature dependent at low temperatures as in Fermi liquids.
It is consistent with the well-known single-band result that SDW will become irrelevant when $U<0$. In the NQR experiment on K$_{2}$Cr$_{3}$As$_{3}$, $1/T_{1}$ exhibits non-integer power law temperature dependence and gives rise to
$1-\frac{U}{2\pi v_F}\sim 0.75$. However, the NQR experiment on Rb$_{2}$Cr$_{3}$As$_{3}$ shows linear temperature dependence at high temperature while critical spin fluctuations appear near to the SC transition temperature $T_c$.
These diversified $1/T_1$ behaviors in K$_{2}$Cr$_{3}$As$_{3}$ and Rb$_{2}$Cr$_{3}$As$_{3}$ imply different effective electron interaction in the two systems. The Rb compound has large unit cell volume than K compound,
resulting in a smaller electron repulsion, which is consistent with larger exponent, $1-\frac{U}{2\pi v_F} \sim 1$ in $1/T_1$ in Rb$_{2}$Cr$_{3}$As$_{3}$.

We next discuss two possible SC ground states in physical parameter regions $0<J<U/3$ and $U/3<J<U/2$.
(1) At $0<J<U/3$, the order parameter $O_{pp}^{20}$ indicates that the SC pairing is spin-singlet and orbital antisymmetric, and the pairing electrons come from the two degenerate $E^{\prime}$ bands.
(2) While for $U/3<J<U/2$, the order parameter $O_{pp}^{23}$ gives rise to spin-triplet ($\left|\uparrow\downarrow\right\rangle+\left|\downarrow\uparrow\right\rangle$) and orbital antisymmetric SC pairing,
and the pairing electrons come from $E^{\prime}$ bands too. This kind of even-parity, spin-triplet and orbital antisymmetric SC pairing was firstly proposed by Dai \textit{et al.} in the context of iron-pnictide \cite{DaiX08}.
Note that the degeneracy of two $E^{\prime}$ bands plays a crucial role in the formation of SC ground states.

The role of the two degenerate $E^{\prime}$ bands can be also seen from the effective Hamiltonian $H_{int}^{B}$ and order parameters for different ground states.
To do this, we consider the situation when the two-fold degeneracy is slightly lifted, for instance, by inter-chain coupling,
In this case, we have $k_{F+1}\neq k_{F-1}$. Introducing $\Delta k_F = k_{F+1}-k_{F-1}$, we can generalize the bosonic interacting Hamiltonian in Eq.~(\ref{Eq:HB-int}) to the expression in Eq.~(\ref{Eq:HB-int2}) in Appendix~\ref{App:HB-int},
where an additional phase factor $2\Delta k_F x$ appears in $g_{2\parallel}^{(1)}$, $g_{2\perp}^{(1)}$ and $g_{1\perp}^{(2)}$ terms. Thus these terms will be suppressed by this phase factor in the integrand. Consequently,
both the spin-triplet SC order parameter $O_{pp}^{20}$ (for $U/3<J<U/2$) and the spin-singlet SC order parameter $O_{pp}^{23}$ (for $0<J<U/3$) will be suppressed and be modulated by the phase factor $2\Delta k_F x$,
indicating a possible FFLO state when $\Delta k_F \neq 0$ \cite{FF,LO}. This is because that both $O_{pp}^{20}$ and $O_{pp}^{23}$ arise from inter-orbital pairing, namely, pairing between $\pm k_{F+1}$ and $\mp k_{F-1}$.

Then we shall investigate how the lifted degeneracy will affect the SDW gound states characterized by order parameters, $O_{ph}^{03}+\frac{\sqrt{3}}{2}O_{ph}^{83}$, $O_{ph}^{43}(O_{ph}^{63})$, and $O_{ph}^{13}$ respectively.
The expression for $O_{ph}^{43}(O_{ph}^{63})$ and $O_{ph}^{13}$ will not change as we turn on $\Delta k_F$. Since SDW instabilites in these states come from the scattering from $\pm k_{F+1}$ to $\mp k_{F-1}$.
However, the order parameter $O_{ph}^{03}+\frac{\sqrt{3}}{2}O_{ph}^{83}$ arising from intra-orbital scattering, say, from $\pm k_{F+1}$ to $\pm k_{F-1}$, will be suppressed and be modulated by the phase factor $2\Delta k_F x$ too.

\begin{figure}
\begin{center}
\includegraphics[width=8.4cm]{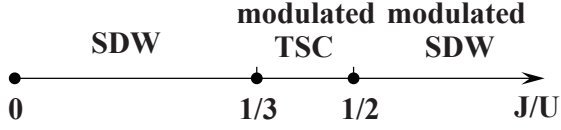}
\caption{Phase diagram for the three-band Hubbard model with lifted degeneracy in $E^{\prime}$ orbitals, $k_{F+1}\neq k_{F-1}$.}
\label{fig:phase_diagram2}
\end{center}
\end{figure}

Thus, we expect that (1) for $0<J<U/3$, the SDW states will win out since the SSC state is suppressed;
(2) for $U/3<J<U/2$, the TSC state will survive and be modulated by a phase factor $2\Delta k_F x$, since the possible competing SDW order ($O_{ph}^{03}+\frac{\sqrt{3}}{2}O_{ph}^{83}$) will be suppressed too;
(3) for the unphysical region $J>U/2$, the SDW state will be modulated by a phase factor $2\Delta k_F x$. The new phase diagram is illustrated in Fig.~\ref{fig:phase_diagram2}.

Finally, we would like to point out that these ordered states will not survive in a single chain due to strong quantum fluctuations,
as stated by Mermin-Wagner-Hohenberg theorm \cite{Mermin-Wagner,Hohenberg}. However, these instabilities will be enhanced at low temperatures, so that small inter-chain couplings will stabilize these ordered states. Moreover, the inter-chain couplings will also determine the spatial pairing symmetry for SC states. Work along this line is under progress.

In summary, we have studied a three-band Hubbard model at incommensurate filling with intra-orbital electron repulsion $U$, inter-orbital electron repulsion $U^{\prime}=U-2J$, and Hund's coupling $J>0$.
With the help of bosonization and RG, we find that the Tomonaga-Luttinger fixed point gives rise to the experimental observed normal state at high temperature.
The ground state instability depends on the ratio $J/U$ and the degeneracy of $E^{\prime}$ bands. When the two $E^{\prime}$ bands are degenerate,
for $0<J<U/3$, the ground state is a spin-singlet SC state; for $U/3<J<U/2$, a spin-triplet SC state is favored;
a SDW state can be achieved in the parameter region $J>U/2$. However, when the two-fold degeneracy of $E^{\prime}$ bands is lifted,
the phase diagram will change. In the physically relevant regions, a SDW state may dominate instead of the spin-singlet SC state when $0<J<U/3$, the spin-triplet SC state is still favored when $U/3<J<U/2$;
but the SC order parameter will be modulated by a spatially varying phase factor $2\Delta k_F x$. Our theoretical results support the existence of a spin-triplet SC state in K$_{2}$Cr$_{3}$As$_{3}$.

\section{acknowledgement}

We would like to thank Guang-Han Cao, Chao Cao, Jian-Hui Dai, Xiao-Yong Feng for helpful discussions, and thank Jun-ichi Okamoto and A. J. Millis for the communications on the two-band situation.
This work is partially supported by National Basic Research Program of China
(No.2014CB921201/2014CB921203), National Key R\&D Program of the MOST of China (No.2016YFA0300202), NSFC (No.11374256/11274269), and the Fundamental Research Funds for the Central Universities in China.

\appendix

\section{Derivation of tree-level RG equations}\label{App:tree-RG}

The tree-level RG equations are given by the scaling dimension $\Delta_k$ for each coupling constant $g_k$. Since $\Delta_k$ is determined by Tomonaga-Luttinger parameters $K_{\mu\nu}$, one finds that
\begin{subequations}\label{Eq:tree-RG1}
\begin{align}
\frac{dg_{1\perp}^{(1)}}{dl} = \left[2-\left(\frac{1}{3}K_{s-1}+\frac{2}{3}K_{s0}+K_{s+1}^{-1}\right)\right]g_{1\perp}^{(1)},
\end{align}
\begin{align}
\frac{dg_{2\parallel}^{(1)}}{dl} = \left[2-\left(K_{c+1}+K_{s+1}\right)\right]g_{2\parallel}^{(1)},
\end{align}
\begin{align}
\frac{dg_{2\perp}^{(1)}}{dl} = \left[2-\left(K_{c+1}+\frac{1}{3}K_{s-1}+\frac{2}{3}K_{s0}\right)\right]g_{2\perp}^{(1)},
\end{align}
\begin{align}
\frac{dg_{3\parallel}^{(1)}}{dl} = \left[2-\left(K_{c+1}^{-1}+K_{s+1}^{-1}\right)\right]g_{3\parallel}^{(1)},
\end{align}
\begin{align}
\frac{dg_{3\perp}^{(1)}}{dl} = \left[2-\left(K_{c+1}^{-1}+\frac{1}{3}K_{s-1}+\frac{2}{3}K_{s0}\right)\right]g_{3\perp}^{(1)},
\end{align}
\begin{align}
\frac{dg_{4\perp}^{(1)}}{dl} = \left[2-\left(K_{c+1}+\frac{1}{3}K_{s-1}+\frac{2}{3}K_{s0}\right)\right]g_{4\perp}^{(1)},
\end{align}
\begin{align}
\frac{dg_{1\perp}^{(2)}}{dl} = \left[2-\left(K_{c+1}+K_{s+1}^{-1}\right)\right]g_{1\perp}^{(2)},
\end{align}
\begin{align}
\frac{dg_{3\perp}^{(2)}}{dl} = \left[2-\left(K_{c+1}^{-1}+K_{s+1}\right)\right]g_{3\perp}^{(2)},
\end{align}
\begin{align}
\frac{df_{1\perp}^{(1)}}{dl} = &\left[2-\left(\frac{1}{4}K_{s+1}+\frac{1}{12}K_{s-1}+\frac{2}{3}K_{s0}+\frac{1}{4}K_{s+1}^{-1}\right.\right. \nonumber \\
                                          &\left.\left.+\frac{3}{4}K_{s-1}^{-1}\right)\right]f_{1\perp}^{(1)},
\end{align}
\begin{align}
\frac{df_{3\parallel}^{(1)}}{dl} = & \left[2-\left(\frac{1}{4}K_{c+1}^{-1}+\frac{3}{4}K_{c-1}^{-1}+\frac{1}{4}K_{s+1}^{-1}+\frac{3}{4}K_{s-1}^{-1}\right)\right] \nonumber\\
                                             & \times f_{3\parallel}^{(1)},
\end{align}
\begin{align}
\frac{df_{3\perp}^{(1)}}{dl} = &\left[2-\left(\frac{1}{4}K_{c+1}^{-1}+\frac{3}{4}K_{c-1}^{-1}+\frac{1}{4}K_{s+1}+\frac{1}{12}K_{s-1}\right.\right. \nonumber \\
                                          &\left.\left.+\frac{2}{3}K_{s0}\right)\right]f_{3\perp}^{(1)},
\end{align}
\begin{align}
\frac{df_{3\perp}^{(2)}}{dl} = &\left[2-\left(\frac{1}{4}K_{c+1}^{-1}+\frac{3}{4}K_{c-1}^{-1}+\frac{1}{4}K_{s+1}+\frac{3}{4}K_{s-1}\right)\right] \nonumber\\
                                         & \times f_{3\perp}^{(2)},
\end{align}
\begin{align}
\frac{dg_{\perp}^{(1)}}{dl}&=\left[2-\left(\frac{4}{3}K_{s-1}+\frac{2}{3}K_{s0}\right)\right]g_{\perp}^{(1)}.
\end{align}
\end{subequations}

To simplify, we introduce the dimensionless coupling constants $x_i$ and $y_i$ as defined in Eq.~(\ref{Eq:xyi}).
In weak coupling, we can expand Tomonaga-Luttinger liquid parameters $K_{\mu\nu}$ to leading order of $x_i$ and $y_i$'s,
\begin{equation}
K_{\mu\nu}=1-y_{\mu\nu},
\end{equation}
where the redefined Tomonaga-Luttinger liquid parameters $y_{\mu\nu}$ are linear combinations of $y_i$'s,
\begin{subequations}\label{Eq:K-y}
\begin{align}
y_{c\pm1}&=\frac{1}{2}\left[\left(y_{4\perp}^{(2)}\right)-\left(y_{2\parallel}^{(2)}+y_{2\perp}^{(2)}\right)\right],  \\
y_{c0}&=\frac{1}{2}\left[\left(y_{4\perp}^{(2)}\right)+2\left(y_{2\parallel}^{(2)}+y_{2\perp}^{(2)}\right)\right],  \\
y_{s\pm1}&=\frac{1}{2}\left[\left(-y_{4\perp}^{(2)}\right)-\left(y_{2\parallel}^{(2)}-y_{2\perp}^{(2)}\right)\right],  \\
y_{s0}&=\frac{1}{2}\left[\left(-y_{4\perp}^{(2)}\right)+2\left(y_{2\parallel}^{(2)}-y_{2\perp}^{(2)}\right)\right].
\end{align}
\end{subequations}
Then the tree-level RG equations Eqs.~(\ref{Eq:tree-RG1}) can be written in terms of $x_i$ and $y_i$'s as in Eqs.~(\ref{Eq:tree-RG2}).

\section{Derivation of one-loop RG equations}\label{App:oneloop-RG}

The full structure constants $C_{ij}^{k}$ in OPE in Eq.~(\ref{Eq:general-RGE}) can be obtained by the fusion of arbitrary two terms in the interaction $H_{int}^B$.
This processes will generate new terms, which are absent in the original microscopic Hamiltonian, until all terms form a closed algebra. Thus we should retain the related terms with zero initial values of coupling constants.
The one-loop RG equations from OPE read,
\begin{subequations}\label{Eq:one-loop}
\begin{align}
\frac{dy_{1\perp}^{\left(1\right)}}{dl}=\left[\left(y_{2\parallel}^{\left(2\right)}-y_{2\perp}^{\left(2\right)}\right)y_{1\perp}^{\left(1\right)}\right]_{1}+\left[-y_{2\perp}^{\left(1\right)}y_{1\perp}^{\left(2\right)}+y_{3\parallel}^{\left(1\right)}y_{3\perp}^{\left(1\right)}\right]_{2},
\end{align}
\begin{align}
\frac{dy_{2\parallel}^{\left(1\right)}}{dl}=\left[-y_{2\parallel}^{\left(2\right)}y_{2\parallel}^{\left(1\right)}\right]_{1}+\left[-y_{2\perp}^{\left(1\right)}y_{4\perp}^{\left(1\right)}\right]_{2},
\end{align}
\begin{align}
\frac{dy_{2\perp}^{\left(1\right)}}{dl}=\left[-y_{2\perp}^{\left(2\right)}y_{2\perp}^{\left(1\right)}\right]_{1}+\left[-y_{1\perp}^{\left(1\right)}y_{1\perp}^{\left(2\right)}-y_{2\parallel}^{\left(1\right)}y_{4\perp}^{\left(1\right)}\right]_{2},
\end{align}
\begin{align}
\frac{dy_{3\parallel}^{\left(1\right)}}{dl}=\left[y_{2\parallel}^{\left(2\right)}y_{3\parallel}^{\left(1\right)}\right]_{1}+\left[y_{1\perp}^{\left(1\right)}y_{3\perp}^{\left(1\right)}\right]_{2},
\end{align}
\begin{align}
\frac{dy_{3\perp}^{\left(1\right)}}{dl}=\left[\left(-y_{4\perp}^{\left(2\right)}+y_{2\parallel}^{\left(2\right)}\right)y_{3\perp}^{\left(1\right)}\right]_{1}+\left[y_{1\perp}^{\left(1\right)}y_{3\parallel}^{\left(1\right)}-y_{4\perp}^{\left(1\right)}y_{3\perp}^{\left(2\right)}\right]_{2},
\end{align}
\begin{align}
\frac{dy_{4\perp}^{\left(1\right)}}{dl}=\left[-y_{2\perp}^{\left(2\right)}y_{4\perp}^{\left(1\right)}\right]_{1}+\left[-y_{2\parallel}^{\left(1\right)}y_{2\perp}^{\left(1\right)}-y_{3\perp}^{\left(1\right)}y_{3\perp}^{\left(2\right)}\right]_{2},
\end{align}
\begin{align}
\frac{dy_{1\perp}^{\left(2\right)}}{dl}=\left[\left(y_{4\perp}^{\left(2\right)}-y_{2\perp}^{\left(2\right)}\right)y_{1\perp}^{\left(2\right)}\right]_{1}+\left[-y_{1\perp}^{\left(1\right)}y_{2\perp}^{\left(1\right)}\right]_{2},
\end{align}
\begin{align}
\frac{dy_{3\perp}^{\left(2\right)}}{dl}=\left[\left(-y_{4\perp}^{\left(2\right)}+y_{2\perp}^{\left(2\right)}\right)y_{3\perp}^{\left(2\right)}\right]_{1}+\left[-y_{3\perp}^{\left(1\right)}y_{4\perp}^{\left(1\right)}\right]_{2},
\end{align}
\begin{align}\label{Eq;decouple_1}
\frac{dx_{1\perp}^{\left(1\right)}}{dl}=\left[\left(y_{2\parallel}^{\left(2\right)}-y_{2\perp}^{\left(2\right)}\right)x_{1\perp}^{\left(1\right)}\right]_{1}+\left[x_{3\parallel}^{\left(1\right)}x_{3\perp}^{\left(1\right)}\right]_{2},
\end{align}
\begin{align}
\frac{dx_{3\parallel}^{\left(1\right)}}{dl}=\left[y_{2\parallel}^{\left(2\right)}x_{3\parallel}^{\left(1\right)}\right]_{1}+\left[x_{1\perp}^{\left(1\right)}x_{3\perp}^{\left(1\right)}\right]_{2},
\end{align}
\begin{align}
\frac{dx_{3\perp}^{\left(1\right)}}{dl}=\left[\left(-y_{4\perp}^{\left(2\right)}+y_{2\parallel}^{\left(2\right)}\right)x_{3\perp}^{\left(1\right)}\right]_{1}+\left[x_{1\perp}^{\left(1\right)}x_{3\parallel}^{\left(1\right)}-y_{\perp}^{\left(1\right)}x_{3\perp}^{\left(2\right)}\right]_{2},
\end{align}
\begin{align}
\frac{dx_{3\perp}^{\left(2\right)}}{dl}=\left[\left(-y_{4\perp}^{\left(2\right)}+y_{2\perp}^{\left(2\right)}\right)x_{3\perp}^{\left(2\right)}\right]_{1}+\left[-y_{\perp}^{\left(1\right)}x_{3\perp}^{\left(1\right)}\right]_{2},
\end{align}
\begin{align}\label{Eq;decouple_2}
\frac{dy_{\perp}^{\left(1\right)}}{dl}=\left[-y_{4\perp}^{\left(2\right)}y_{\perp}^{\left(1\right)}\right]_{1}+\left[-x_{3\perp}^{\left(1\right)}x_{3\perp}^{\left(2\right)}\right]_{2}.
\end{align}
\end{subequations}
To distinguish the tree-level and one-loop contribution in the full one-loop RG equations, we use subscript $\left[\cdots\right]_{1}$ to denote the tree-level terms and $\left[\cdots\right]_{2}$ to denote the one-loop terms.

As mentioned in Section~\ref{sec:tree-RG} in the main text, we should restore the spin rotational symmetry by imposing the following constraints on coupling constants,
\begin{subequations}\label{Eq:SU2}
\begin{align}
g_{1\parallel}^{(1)}-g_{2\parallel}^{(2)}=g_{1\perp}^{(1)}-g_{2\perp}^{(2)},  \\
g_{2\parallel}^{(1)}-g_{1\parallel}^{(2)}=g_{2\perp}^{(1)}-g_{1\perp}^{(2)},  \\
g_{3\parallel}^{(1)}-g_{3\parallel}^{(2)}=g_{3\perp}^{(1)}-g_{3\perp}^{(2)},  \\
g_{4\parallel}^{(1)}-g_{4\parallel}^{(2)}=g_{4\perp}^{(1)}-g_{4\perp}^{(2)},  \\
f_{1\parallel}^{(1)}-f_{2\parallel}^{(2)}=f_{1\perp}^{(1)}-f_{2\perp}^{(2)},  \\
f_{2\parallel}^{(1)}-f_{1\parallel}^{(2)}=f_{2\perp}^{(1)}-f_{1\perp}^{(2)},  \\
f_{3\parallel}^{(1)}-f_{3\parallel}^{(2)}=f_{3\perp}^{(1)}-f_{3\perp}^{(2)},  \\
g_{\parallel}^{(1)}-g_{\parallel}^{(2)}=g_{\perp}^{(1)}-g_{\perp}^{(2)}.
\end{align}
\end{subequations}
which will further simplify our analysis. In addition to these spin $SU(2)$ constraints, we also have 
\begin{subequations}\label{constraint_added}
\begin{eqnarray}
g_{4\perp}^{(2)} & = & g_{\perp}^{(2)},\\
g_{2\parallel}^{(2)} & = & f_{2\parallel}^{(2)},\\
g_{2\perp}^{(2)} & = & f_{2\perp}^{(2)}.
\end{eqnarray}
\end{subequations}
from the microscopic Hamiltonian.
By these relations, the coupling constants $y_{4\perp}^{\left(2\right)}$, $y_{2\parallel}^{\left(2\right)}$ and $y_{2\perp}^{\left(2\right)}$ can substitute by $y_{\perp}^{\left(2\right)}$, $x_{2\parallel}^{\left(2\right)}$ and $x_{2\perp}^{\left(2\right)}$ 
in Eqs.~(\ref{Eq:one-loop}). Thus the 13 RG equations (\ref{Eq:one-loop}) can be decoupled into two sets, one consists of 8 equations and the other consists of 5 equations, as the following,
\begin{subequations}\label{Eq:one-loop2}
\begin{align}
\frac{dy_{1\perp}^{(1)}}{dl}  = & -\left(y_{1\perp}^{(1)}\right)^{2}-y_{2\perp}^{(1)}y_{1\perp}^{(2)}+y_{3\parallel}^{(1)}y_{3\perp}^{(1)},
\end{align}
\begin{align}
\frac{dy_{2\parallel}^{(1)}}{dl}  = & -\frac{1}{2}\left(y_{2\parallel}^{(2)}+y_{2\perp}^{(2)}-y_{1\perp}^{(1)}\right)y_{2\parallel}^{(1)}-y_{2\perp}^{(1)}y_{4\perp}^{(1)},
\end{align}
\begin{align}
\frac{dy_{2\perp}^{(1)}}{dl}  = & -\frac{1}{2}\left(y_{2\parallel}^{(2)}+y_{2\perp}^{(2)}+y_{1\perp}^{(1)}\right)y_{2\perp}^{(1)}-y_{1\perp}^{(1)}y_{1\perp}^{(2)}\nonumber\\
                                & -y_{2\parallel}^{(1)}y_{4\perp}^{(1)},
\end{align}
\begin{align}
\frac{dy_{3\parallel}^{(1)}}{dl}  = & \frac{1}{2}\left(y_{2\parallel}^{(2)}+y_{2\perp}^{(2)}-y_{1\perp}^{(1)}\right)y_{3\parallel}^{(1)}+y_{1\perp}^{(1)}y_{3\perp}^{(1)},
\end{align}
\begin{align}
\frac{dy_{3\perp}^{(1)}}{dl}  = & \left(-y_{4\perp}^{(1)}+\frac{1}{2}\left(y_{2\parallel}^{(2)}+y_{2\perp}^{(2)}-y_{1\perp}^{(1)}\right)\right)y_{3\perp}^{(1)}\nonumber\\
                                & +y_{1\perp}^{(1)}y_{3\parallel}^{(1)}-y_{4\perp}^{(1)}y_{3\perp}^{(2)},
\end{align}
\begin{align}
\frac{dy_{4\perp}^{(1)}}{dl}  = & -\frac{1}{2}\left(y_{2\parallel}^{(2)}+y_{2\perp}^{(2)}-y_{1\perp}^{(1)}\right)y_{4\perp}^{(1)}-y_{2\parallel}^{(1)}y_{2\perp}^{(1)}\nonumber\\
                                & -y_{3\perp}^{(1)}y_{3\perp}^{(2)},
\end{align}
\begin{align}
\frac{dy_{1\perp}^{(2)}}{dl}  = & \left(y_{4\perp}^{(1)}-\frac{1}{2}\left(y_{2\parallel}^{(2)}+y_{2\perp}^{(2)}+y_{1\perp}^{(1)}\right)\right)y_{1\perp}^{(2)}\nonumber\\
                                & -y_{1\perp}^{(1)}y_{2\perp}^{(1)},
\end{align}
\begin{align}
\frac{dy_{3\perp}^{(2)}}{dl}  = & \left(-y_{4\perp}^{(1)}+\frac{1}{2}\left(y_{2\parallel}^{(2)}+y_{2\perp}^{(2)}+y_{1\perp}^{(1)}\right)\right)y_{3\perp}^{(2)}\nonumber\\
                                & -y_{3\perp}^{(1)}y_{4\perp}^{(1)},
\end{align}
\begin{align}
\frac{dx_{1\perp}^{(1)}}{dl}  = & -\left(x_{1\perp}^{(1)}\right)^{2}+x_{3\parallel}^{(1)}x_{3\perp}^{(1)},
\end{align}
\begin{align}
\frac{dx_{3\parallel}^{(1)}}{dl}  = & \frac{1}{2}\left(x_{2\parallel}^{(2)}+x_{2\perp}^{(2)}-x_{1\perp}^{(1)}\right)x_{3\parallel}^{(1)}+x_{1\perp}^{(1)}x_{3\perp}^{(1)},
\end{align}
\begin{align}
\frac{dx_{3\perp}^{(1)}}{dl}  = & \left(-y_{\perp}^{(1)}+\frac{1}{2}\left(x_{2\parallel}^{(2)}+x_{2\perp}^{(2)}-x_{1\perp}^{(1)}\right)\right)x_{3\perp}^{(1)}\nonumber\\
                                & +x_{1\perp}^{(1)}x_{3\parallel}^{(1)}-y_{\perp}^{\left(1\right)}x_{3\perp}^{\left(2\right)},
\end{align}
\begin{align}\
\frac{dx_{3\perp}^{(2)}}{dl}  = & \left(-y_{\perp}^{(1)}+\frac{1}{2}\left(x_{2\parallel}^{(2)}+x_{2\perp}^{(2)}+x_{1\perp}^{(1)}\right)\right)x_{3\perp}^{(2)}-y_{\perp}^{\left(1\right)}x_{3\perp}^{\left(1\right)},
\end{align}
\begin{align}
\frac{dy_{\perp}^{(1)}}{dl}  = & -\left(y_{\perp}^{(1)}\right)^{2}-x_{3\perp}^{\left(1\right)}x_{3\perp}^{\left(2\right)}.
\end{align}
\end{subequations}

Note that the variables $y_{2\parallel}^{(2)}+y_{2\perp}^{(2)}$ and $x_{2\parallel}^{(2)}+x_{2\perp}^{(2)}$ do not flow as they do not appear in $H_{int}^B$.
In perturbation, we can drop these small variables to obtain closed one-loop RG equations.
By keeping all the relevant terms, we have Eqs.~(\ref{Eq:one-loopRGE}).

\section{Spin-lattice relaxation rate for a three-band Tomonaga-Luttinger liquid}\label{App:T1}

The spin-lattice relaxation rate $1/T_{1}$ in a NMR experiment detects the local spin correlation function
\begin{equation}
\frac{1}{T_{1}}=A_{f}^{2}T\sum_{q}\frac{Im\chi\left(q,\omega\right)}{\omega},
\end{equation}
where $A_f$ is the hyperfine coupling constant. The spin correlation function in a multi-band system is defined as
\begin{equation}
\chi^{\alpha\beta}\left(x,\tau\right)=\sum_{m}\left\langle S_{m}^{\alpha}\left(x,\tau\right)S_{m}^{\beta}\left(0,0\right)\right\rangle
\end{equation}
where $S_{m}^{\alpha}=\frac{1}{2}\sum_{ss^{\prime}}c_{ms}^{\dagger}\sigma_{ss^{\prime}}^{\alpha}c_{ms^{\prime}}$ is the spin operator in the $m$-th band and $\alpha,\beta=x,y,z$.

For a system where spin rotational symmetry is respected, $\chi^{\alpha\beta}=\chi\delta_{\alpha\beta}$. We find that the dominant contribution to $\chi$ come from $q\sim 0$ and $q\sim 2k_F$ components.
Summing over all the bands, we have the following form of the temperature dependent spin-lattice relaxation rate,
\begin{eqnarray}\label{Eq:T1-1}
\frac{1}{T_{1}} & \propto & A_1 T\nonumber \\
 & + & A_2 T^{\frac{1}{2}\left[\left(K_{c+1}+\frac{1}{3}K_{c-1}+\frac{2}{3}K_{c0}\right)+\left(K_{s+1}+\frac{1}{3}K_{s-1}+\frac{2}{3}K_{s0}\right)\right]-1}\nonumber \\
 & + & A_3 T^{\frac{1}{2}\left[\left(\frac{4}{3}K_{c-1}+\frac{2}{3}K_{c0}\right)+\left(\frac{4}{3}K_{s-1}+\frac{2}{3}K_{s0}\right)\right]-1}.
\end{eqnarray}
The first linearly temperature dependent term follows Korringa law as in Fermi liquids. The second term comes from the two degenerate bands $m=\pm1$ and both have the same exponent. The third term comes from the non-degenerate band $m=0$.
In the non-interacting limit, $K_{\mu\nu}=1$, thus all the three terms become linearly temperature dependent as in Fermi liquids.

With the help of spin rotational symmetry, we can further simplify the expression in Eq.~(\ref{Eq:T1-1}). Considering the spin correlation function $\chi^{\alpha\beta}=\chi\delta_{\alpha\beta}$ in the TLL fixed point,
the spin $SU(2)$ symmetry will imposes the following constraints to Tomonaga-Luttinger parameters,
\begin{equation}
K_{sm}=1.
\end{equation}
Meanwhile, we have $K_{c+1}=K_{c-1}$ in our model. In weak coupling, we can expand Tomonaga-Luttinger parameters as $K_{\mu\nu}=1-y_{\mu\nu}$. Then the spin-lattice relaxation rate can be simplified as
\begin{equation}
\frac{1}{T_{1}}\propto A~T + B~T^{1-\frac{y_{4\perp}^{(2)}}{2}},
\end{equation}
where the coupling constant $y_{4\perp}^{(2)}$ is chosen as its initial value $y_{4\perp}^{(2)}=U/\pi v_{F}$.
When the short ranged Coulomb repulsion govern the system, $U$ is positive, the dominant contribution will come from the second term at low temperatures.
The spin-lattice relaxation rate $1/T_{1}$ will exhibit non-integer power law temperature dependence.
However, $U$ may become negative effectively, e.g., when electron-phonon interaction dominates. In this case, $1/T_{1}$ will become linearly temperature dependent at low temperatures as in Fermi liquids.
It is consistent with the common sense that SDW terms will become irrelevant when $U<0$.

\section{Interacting Hamiltonian $H_{int}^B$ for $k_{F+1}\neq k_{F-1}$}\label{App:HB-int}

When the degeneracy of the two $E^{\prime}$ bands is lifted, $k_{F+1}\neq k_{F-1}$. From Eq.~(\ref{Eq:diagonalize-HB0}), we find that the terms containing phase $\tilde{\phi}_{c+1}\propto \phi_{c+1}-\phi_{c-1}$ should change with a phase factor
$\Delta k_F x$, resulting in the interacting Hamiltonian $H_{int}^B$ for $k_{F+1}\neq k_{F-1}$ as follows,
\begin{widetext}
\begin{eqnarray}\label{Eq:HB-int2}
H_{int}^{B}= & - & g_{1\perp}^{\left(1\right)}\frac{4}{\left(2\pi a\right)^{2}}\int dx\cos\left(\frac{2}{\sqrt{3}}\tilde{\phi}_{s-1}+\frac{4}{\sqrt{6}}\tilde{\phi}_{s0}\right)\cos\left(2\tilde{\theta}_{s+1}\right)\nonumber \\
 & + & \underline{g_{2\parallel}^{\left(1\right)}\frac{4}{\left(2\pi a\right)^{2}}\int dx\cos\left(2\Delta k_F x+2\tilde{\phi}_{c+1}\right)\cos\left(2\tilde{\phi}_{s+1}\right)}\nonumber \\
 & + & \underline{g_{2\perp}^{\left(1\right)}\frac{4}{\left(2\pi a\right)^{2}}\int dx\cos\left(2\Delta k_F x+2\tilde{\phi}_{c+1}\right)\cos\left(\frac{2}{\sqrt{3}}\tilde{\phi}_{s-1}+\frac{4}{\sqrt{6}}\tilde{\phi}_{s0}\right)}\nonumber \\
 & + & g_{3\parallel}^{\left(1\right)}\frac{4}{\left(2\pi a\right)^{2}}\int dx\cos\left(2\tilde{\theta}_{c+1}\right)\cos\left(2\tilde{\theta}_{s+1}\right)\nonumber \\
 & + & g_{3\perp}^{\left(1\right)}\frac{4}{\left(2\pi a\right)^{2}}\int dx\cos\left(2\tilde{\theta}_{c+1}\right)\cos\left(\frac{2}{\sqrt{3}}\tilde{\phi}_{s-1}+\frac{4}{\sqrt{6}}\tilde{\phi}_{s0}\right)\nonumber \\
 & + & g_{4\perp}^{\left(1\right)}\frac{4}{\left(2\pi a\right)^{2}}\int dx\cos\left(2\tilde{\phi}_{s+1}\right)\cos\left(\frac{2}{\sqrt{3}}\tilde{\phi}_{s-1}+\frac{4}{\sqrt{6}}\tilde{\phi}_{s0}\right)\nonumber \\
 & - & \underline{g_{1\perp}^{\left(2\right)}\frac{4}{\left(2\pi a\right)^{2}}\int dx\cos\left(2\Delta k_F x+2\tilde{\phi}_{c+1}\right)\cos\left(2\tilde{\theta}_{s+1}\right)}\nonumber \\
 & + & g_{3\perp}^{\left(2\right)}\frac{4}{\left(2\pi a\right)^{2}}\int dx\cos\left(2\tilde{\theta}_{c+1}\right)\cos\left(2\tilde{\phi}_{s+1}\right)\nonumber \\
 & - & f_{1\perp}^{\left(1\right)}\frac{8}{\left(2\pi a\right)^{2}}\int dx\left[\cos\tilde{\phi}_{s+1}\cos\left(-\frac{1}{\sqrt{3}}\tilde{\phi}_{s-1}+\frac{4}{\sqrt{6}}\tilde{\phi}_{s0}\right)\cos\tilde{\theta}_{s+1}\cos\sqrt{3}\tilde{\theta}_{s-1}+\left(\cos\rightarrow\sin\right)\right]\nonumber \\
 & + & f_{3\parallel}^{\left(1\right)}\frac{8}{\left(2\pi a\right)^{2}}\int dx\left[\cos\tilde{\theta}_{c+1}\cos\sqrt{3}\tilde{\theta}_{c-1}\cos\tilde{\theta}_{s+1}\cos\sqrt{3}\tilde{\theta}_{s-1}+\left(\cos\rightarrow\sin\right)\right]\nonumber \\
 & + & f_{3\perp}^{\left(1\right)}\frac{8}{\left(2\pi a\right)^{2}}\int dx\left[\cos\tilde{\theta}_{c+1}\cos\sqrt{3}\tilde{\theta}_{c-1}\cos\tilde{\phi}_{s+1}\cos\left(-\frac{1}{\sqrt{3}}\tilde{\phi}_{s-1}+\frac{4}{\sqrt{6}}\tilde{\phi}_{s0}\right)+\left(\cos\rightarrow\sin\right)\right]\nonumber \\
 & + & f_{3\perp}^{\left(2\right)}\frac{8}{\left(2\pi a\right)^{2}}\int dx\left[\cos\tilde{\theta}_{c+1}\cos\sqrt{3}\tilde{\theta}_{c-1}\cos\tilde{\phi}_{s+1}\cos\sqrt{3}\tilde{\phi}_{s-1}+\left(\cos\rightarrow\sin\right)\right]\nonumber \\
 & + & g_{\,\,\,\perp}^{\left(1\right)}\frac{2}{\left(2\pi a\right)^{2}}\int dx\cos\left(-\frac{4}{\sqrt{3}}\tilde{\phi}_{s-1}+\frac{4}{\sqrt{6}}\tilde{\phi}_{s0}\right),
\end{eqnarray}
where $\Delta k_F = k_{F+1}-k_{F-1}$.
\end{widetext}

\section{Non-perturbative solutions}\label{App:dual}

Indeed, there exist other solutions to the one-loop RG equations \eqref{Eq:one-loopRGE} besides the perturbative solutions discussed in Section \ref{sec:oneloop-RG}.
One example is given by the following fixed points satisfying Eq.~(\ref{Eq:fixed point}), 
\begin{subequations}\label{Eq:fixed 8}
\begin{eqnarray}
y_{1\perp}^{\left(1\right)}=y_{4\perp}^{\left(1\right)} & = & 0,\\
-y_{2\perp}^{\left(1\right)}y_{1\perp}^{\left(2\right)}+y_{3\parallel}^{\left(1\right)}y_{3\perp}^{\left(1\right)} & = & 0,\\
y_{2\parallel}^{\left(1\right)}y_{2\perp}^{\left(1\right)}+y_{3\perp}^{\left(1\right)}y_{3\perp}^{\left(2\right)} & = & 0,
\end{eqnarray}
\end{subequations}
and
\begin{subequations}\label{Eq:fixed 5}
\begin{eqnarray}
y_{\perp}^{\left(1\right)}=x_{1\perp}^{\left(1\right)} & = & 0,\\
x_{3\parallel}^{\left(1\right)}x_{3\perp}^{\left(1\right)}=x_{3\perp}^{\left(1\right)}x_{3\perp}^{\left(2\right)} & = & 0.
\end{eqnarray}
\end{subequations}
The solution gives rise to five or six dimensional manifold. It is easy to see that these fixed points are critical points except the trivial TLL fixed point $\vec{y}^{*}=0$.
This can be examined through the $W$ matrix defined in Eq.~\eqref{eq:defW}.

As mentioned in Section \ref{sec:oneloop-RG}, the 13 running coupling constants form two separated close sets in Eqs.~(\ref{Eq:one-loopRGE}), 
$\{y_{1\perp}^{(1)}, y_{2\parallel}^{(1)}, y_{2\perp}^{(1)}, y_{3\parallel}^{(1)}, y_{3\perp}^{(1)}, y_{4\perp}^{(1)}, y_{1\perp}^{(2)}, y_{3\perp}^{(2)}\}$ 
and $\{x_{1\perp}^{(1)}, x_{3\parallel}^{(1)}, x_{3\perp}^{(1)},x_{3\perp}^{(2)},y_{\perp}^{(1)} \}$.
So that the $W$ matrix is diagonal block as in Eq.~\eqref{Eq:W12}, and the generic forms of $W_1$ and $W_2$ at a fixed point $\vec{y}=\vec{y}^{*}$ are given as follows,
\begin{widetext}
\begin{equation}\label{eq:gW1}
W_1=\left(\begin{array}{cccccccc}
-2y_{1\perp}^{(1)*} & 0 & -y_{1\perp}^{(2)*} & y_{3\perp}^{(1)*} & y_{3\parallel}^{(1)*} & 0 & -y_{2\perp}^{(1)*} & 0 \\
\frac{1}{2}y_{2\parallel}^{(1)*} & \frac{1}{2}y_{1\perp}^{(1)*} & -y_{4\perp}^{(1)*} & 0 & 0 & -y_{2\perp}^{(1)*} & 0 & 0 \\
-\frac{1}{2}y_{2\perp}^{(1)*}-y_{1\perp}^{(2)*} & -y_{4\perp}^{(1)*} & -\frac{1}{2}y_{1\perp}^{(1)*} & 0 & 0 & -y_{2\parallel}^{(1)*} & -y_{1\perp}^{(1)*} & 0 \\
-\frac{1}{2}y_{3\parallel}^{(1)*}+y_{3\perp}^{(1)*} & 0 & 0 & -\frac{1}{2}y_{1\perp}^{(1)*} & y_{1\perp}^{(1)*} & 0 & 0 & 0 \\
-\frac{1}{2}y_{3\perp}^{(1)*}+y_{3\parallel}^{(1)*} & 0 & 0 & y_{1\perp}^{(1)*} & -y_{4\perp}^{(1)*}-\frac{1}{2}y_{1\perp}^{(1)*} & -y_{3\perp}^{(1)*}-y_{3\perp}^{(2)*} & 0 & -y_{4\perp}^{(1)*} \\
\frac{1}{2}y_{4\perp}^{(1)*} & -y_{2\perp}^{(1)*} & -y_{2\parallel}^{(1)*} & 0 & -y_{3\perp}^{(2)*} & \frac{1}{2}y_{1\perp}^{(1)*} & 0 & -y_{3\perp}^{(1)*} \\
-\frac{1}{2}y_{1\perp}^{(2)*}-y_{2\perp}^{(1)*} & 0 & -y_{1\perp}^{(1)*} & 0 & 0 & y_{1\perp}^{(2)*} & y_{4\perp}^{(1)*}-\frac{1}{2}y_{1\perp}^{(1)*} & 0 \\
\frac{1}{2}y_{3\perp}^{(2)*} & 0 & 0 & 0 & -y_{4\perp}^{(1)*} & -y_{3\perp}^{(1)*}-y_{3\perp}^{(2)*} & 0 & -y_{4\perp}^{(1)*}+\frac{1}{2}y_{1\perp}^{(1)*}
\end{array}\right),
\end{equation}
and
\begin{equation}\label{eq:gW2}
W_2=\left(\begin{array}{ccccc}
-2x_{1\perp}^{(1)*} & x_{3\perp}^{(1)*} & x_{3\parallel}^{(1)*} & 0 & 0 \\
-\frac{1}{2}x_{3\parallel}^{(1)*}+x_{3\perp}^{(1)*} & -\frac{1}{2}x_{1\perp}^{(1)*} & x_{1\perp}^{(1)*} & 0 & 0 \\
-\frac{1}{2}x_{3\perp}^{(1)*}+x_{3\parallel}^{(1)*} & x_{1\perp}^{(1)*} & -y_{\perp}^{(1)*}-\frac{1}{2}x_{1\perp}^{(1)*} &  -y_{\perp}^{(1)*} &  -x_{3\perp}^{(1)*}-x_{3\perp}^{(2)*}\\
\frac{1}{2}x_{3\perp}^{(2)*} & 0 & -y_{\perp}^{(1)*} & -y_{\perp}^{(1)*}+\frac{1}{2}x_{1\perp}^{(1)*} & -x_{3\perp}^{(1)*}-x_{3\perp}^{(2)*} \\
0 & 0 & -x_{3\perp}^{(2)*} & -x_{3\perp}^{(1)*} & -2y_{\perp}^{(1)*}
\end{array}\right).
\end{equation}
\end{widetext}

On the 5D or 6D fixed hypersurface determined by Eqs. \eqref{Eq:fixed 8} and \eqref{Eq:fixed 5}, there exist both relevant and irrelevant scaling fields at each nontrivial fixed point ($\vec{y}*\neq 0$).
For instance, we consider the $W_2$ matrix and the solution,
\begin{subequations}
\begin{eqnarray}
y_{\perp}^{(1)} =x_{1\perp}^{(1)}= x_{3\perp}^{(1)} & = &  0,\\
x_{3\parallel}^{(1)}& = & x_{3\parallel}^{(1)*}, \\
x_{3\perp}^{(2)}& = & x_{3\perp}^{(2)*}.
\end{eqnarray}
\end{subequations}
The $W_2$ matrix has two nonzero eigenvalues with opposite signs, $\pm \sqrt{\left(x_{3\parallel}^{(1)*}\right)^2+\left(x_{3\perp}^{(2)*}\right)^2}$.
Similarly, the solution
\begin{subequations} 
\begin{eqnarray}
y_{\perp}^{(1)} =x_{1\perp}^{(1)}= x_{3\parallel}^{(1)}= x_{3\perp}^{(2)} & = &  0,\\
 x_{3\perp}^{(1)} & = & x_{3\perp}^{(1)*},
\end{eqnarray}
\end{subequations}
also gives rise to eigenvalues $\pm x_{3\perp}^{(1)*}$ for matrix $W_2$.

The 8 variables $\{y_{1\perp}^{(1)}, y_{2\parallel}^{(1)}, y_{2\perp}^{(1)}, y_{3\parallel}^{(1)}, y_{3\perp}^{(1)}, y_{4\perp}^{(1)}, y_{1\perp}^{(2)}, y_{3\perp}^{(2)}\}$ describes the coupling within the two degenerate $E^{\prime}$ bands.
Within this subspace of the total 13-dimensional parameter space, there exist a duality symmetry on the critical hypersurface.
If one exchanges the fields,
\begin{subequations}
\begin{eqnarray}
\tilde{\phi}_{c+1} & \leftrightarrow & \tilde{\theta}_{c+1},\\
\tilde{\phi}_{s+1} & \leftrightarrow & \tilde{\theta}_{s+1},
\end{eqnarray}
\end{subequations}
the forms of Hamiltonian \eqref{Eq:H0B} and \eqref{Eq:HB-int} do not change. So that the particular choice of coupling constants on the fixed hypersuface given by Eqs.~\eqref{Eq:fixed 8} and \eqref{Eq:fixed 5},
\begin{subequations}
\begin{eqnarray}
y_{2\parallel}^{(1)}& = & y_{3\parallel}^{(1)},\\
y_{2\perp}^{(1)} & = & y_{3\perp}^{(1)}, \\
y_{1\perp}^{(2)} & = & -y_{3\perp}^{(2)}, \\
y_{1\perp}^{(1)} & = &  y_{4\perp}^{(1)} = 0, \\
x_{1\perp}^{(1)} & = & x_{3\parallel}^{(1)} = x_{3\perp}^{(1)} = x_{3\perp}^{(2)} = y_{\perp}^{(1)}  =  0,
\end{eqnarray}
\end{subequations}
and $K_{c+1}=K_{s+1}=1$ will give rise to self-dual fixed points. However, the coupling between the $E^{\prime}$ bands and the $A_{1}^{\prime}$ band, 
with nonzero $x_{3\parallel}^{(1)}, x_{3\perp}^{(1)}, x_{3\perp}^{(2)}$ will spoil such a duality symmetry.

\section{Numerical study of one-loop RG equations}\label{App:NRG}

\begin{figure}[hptb]
\begin{centering}
\includegraphics[width=8.0cm]{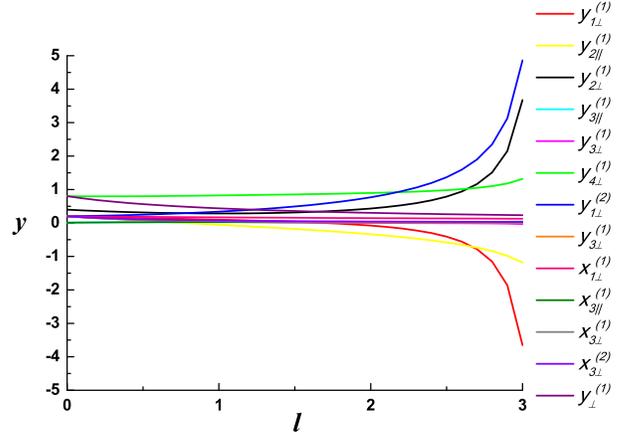}
\caption{(Color online) The RG flows starting from bare values given in Eq.~\eqref{eq:NRGbare} with $\frac{U}{\pi v_{F}}=0.8$ and $\frac{J}{\pi v_{F}}=0.2$.}
\label{fig:NRG_flow_bare}
\end{centering}
\end{figure}

We solve the one-loop RG equations \eqref{Eq:one-loopRGE} numerically and the calculation is carried out in two stages. Firstly, begin with the initial values given by Eq.~\eqref{Eq:couplings},
\begin{equation}\label{eq:NRGbare}
\renewcommand\arraystretch{1.8}
\begin{array}{lll}
y_{1\perp}^{\left(1\right)} = \frac{J}{\pi v_F},  & y_{2\parallel}^{\left(1\right)}  =  \frac{U-3J}{\pi v_F}, & y_{2\perp}^{\left(1\right)} = \frac{U-2J}{\pi v_F},\\
y_{3\parallel}^{\left(1\right)}  = 0, & y_{3\perp}^{\left(1\right)} = \frac{J}{\pi v_F}, & y_{4\perp}^{\left(1\right)}  =  \frac{U}{\pi v_F}, \\
y_{1\perp}^{\left(2\right)} = \frac{J}{\pi v_F}, & y_{3\perp}^{\left(2\right)}  =  \frac{J}{\pi v_F}, & \\
x_{1\perp}^{\left(1\right)} = \frac{J}{\pi v_F}, & x_{3\parallel}^{\left(1\right)}  =  0, & x_{3\perp}^{\left(1\right)} = \frac{J}{\pi v_F},\\
x_{3\perp}^{\left(2\right)}  =  \frac{J}{\pi v_F}, & y_{\perp}^{\left(1\right)} =  \frac{U}{\pi v_F}, & 
\end{array}
\renewcommand\arraystretch{1}
\end{equation}
we find that the running coupling constant $y_{1\perp}^{\left(2\right)}$ always approaches the order of $O(1)$ first,
which is consistent with analytic analysis in Section \ref{sec:tree-RG} and \ref{sec:oneloop-RG} that the most relevant coupling constant is $y_{1\perp}^{\left(2\right)}$.
For instance, the RG flows starting from these bare values given in Eq.~\eqref{eq:NRGbare} with $\frac{U}{\pi v_{F}}=0.8$ and $\frac{J}{\pi v_{F}}=0.2$ are plotted in Fig.~\ref{fig:NRG_flow_bare}.

\begin{figure}[hptb]
\begin{centering}
\includegraphics[width=8.0cm]{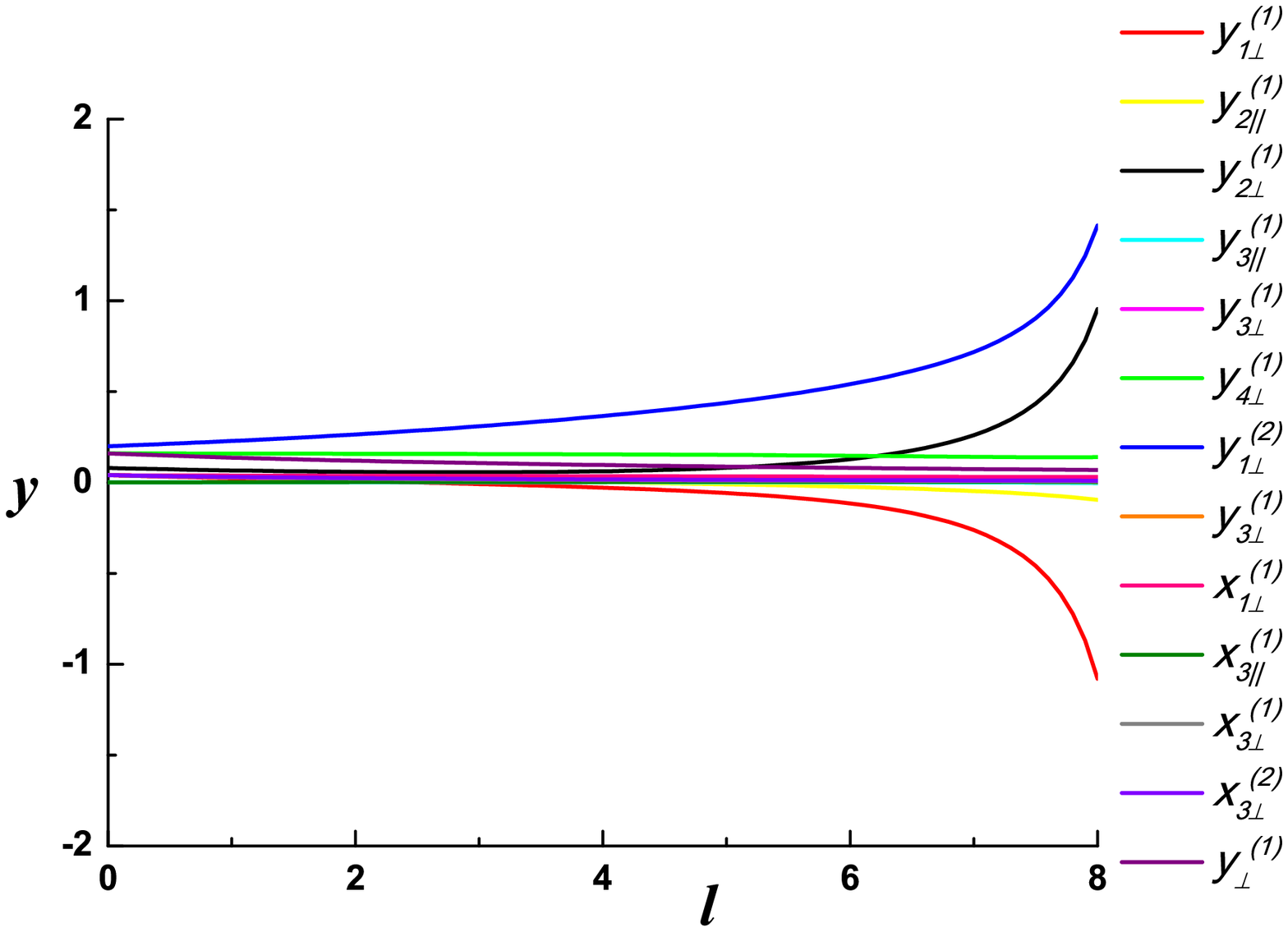}
\caption{(Color online) The RG flows starting from initial values given in Eq.~\eqref{eq:NRGeps} with $\frac{U}{\pi v_{F}}=0.8$ and $\frac{J}{\pi v_{F}}=0.2$, satisfying that $J<U/3$. 
The dominant phase is a spin singlet superconductor, which is given by the order parameter $O_{pp}^{20}$.}
\label{fig:NRG_flow1}
\end{centering}
\end{figure}

\begin{figure}[hptb]
\begin{centering}
\includegraphics[width=8.0cm]{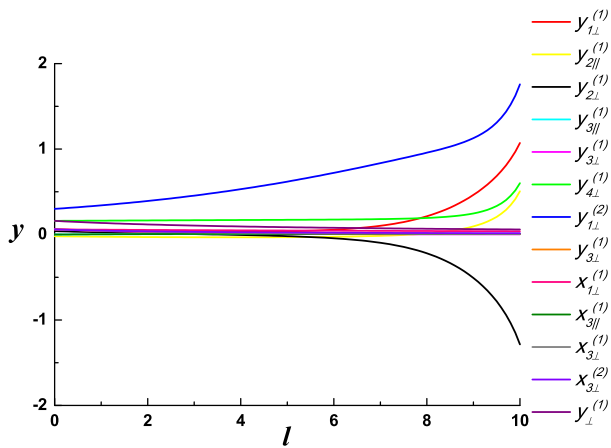}
\caption{(Color online) The RG flows starting from initial values given in Eq.~\eqref{eq:NRGeps} with $\frac{U}{\pi v_{F}}=0.8$ and $\frac{J}{\pi v_{F}}=0.3$, satisfying that $J>U/3$.
The dominant phase is a spin triplet superconductor, which is given by the order parameter $O_{pp}^{23}$.}
\label{fig:NRG_flow2}
\end{centering}
\end{figure}

However, there exist several different ordered states associated with the dominant $y_{1\perp}^{\left(2\right)}$ term. In order to distinguish these competing orders, we will suppress
the bare values by a factor $\varepsilon<1$ except the most relevant one  $y_{1\perp}^{\left(2\right)}$, and choose the initial values as follows,
\begin{equation}\label{eq:NRGeps}
\renewcommand\arraystretch{1.8}
\begin{array}{lll}
y_{1\perp}^{\left(1\right)} = \frac{J}{\pi v_F}\varepsilon,  & y_{2\parallel}^{\left(1\right)}  =  \frac{U-3J}{\pi v_F}\varepsilon, & y_{2\perp}^{\left(1\right)} = \frac{U-2J}{\pi v_F}\varepsilon,\\
y_{3\parallel}^{\left(1\right)}  = 10^{-4}, & y_{3\perp}^{\left(1\right)} = \frac{J}{\pi v_F}\varepsilon, & y_{4\perp}^{\left(1\right)}  =  \frac{U}{\pi v_F}\varepsilon, \\
y_{1\perp}^{\left(2\right)} = \frac{J}{\pi v_F}, & y_{3\perp}^{\left(2\right)}  =  \frac{J}{\pi v_F}\varepsilon, & \\
x_{1\perp}^{\left(1\right)} = \frac{J}{\pi v_F}\varepsilon, & x_{3\parallel}^{\left(1\right)}  =  10^{-4}, & x_{3\perp}^{\left(1\right)} = \frac{J}{\pi v_F}\varepsilon,\\
x_{3\perp}^{\left(2\right)}  =  \frac{J}{\pi v_F}\varepsilon, & y_{\perp}^{\left(1\right)} =  \frac{U}{\pi v_F}\varepsilon. & 
\end{array}
\renewcommand\arraystretch{1}
\end{equation}
There are two distinct situations for different $J/U$ values. (1) When $J/U<1/3$, an example is shown in Fig.~\ref{fig:NRG_flow1}, with $\frac{U}{\pi v_{F}}=0.8$ and $\frac{J}{\pi v_{F}}=0.2$,
the leading terms are $y_{1\perp}^{\left(1\right)}<0$, $y_{2\perp}^{\left(1\right)}>0$ and $y_{1\perp}^{\left(2\right)}>0$. 
These values of coupling constants will lock the bosonic fields at the saddle point with nonzero order parameter $O_{pp}^{20}$. 
Thus we obtain the spin-singlet superconducting (SSC) phase. 
(2) When $J/U>1/3$, as demonstrated in Fig.~\ref{fig:NRG_flow2}, with $\frac{U}{\pi v_{F}}=0.8$ and $\frac{J}{\pi v_{F}}=0.3$, 
the leading terms now become $y_{1\perp}^{\left(1\right)}>0$, $y_{2\perp}^{\left(1\right)}<0$ and $y_{1\perp}^{\left(2\right)}>0$.
By the same method, we obtain the spin-triplet superconducting (TSC) ground state with order parameter $O_{pp}^{23}$. 
We note these three coupling constants $y_{1\perp}^{\left(1\right)}$, $y_{2\perp}^{\left(1\right)}$ and $y_{1\perp}^{\left(2\right)}$ form a closed OPE,
which will become relevant simultaneously. 
The phase boundary determined numerically is about $J/U=1/3$, which is consistent with the $W$-matrix analysis of RG equations as carried in Section \ref{sec:oneloop-RG}.

\end{document}